 \documentclass[final,5p,times,twocolumn]{elsarticle}

\usepackage{amssymb}
\usepackage{url}

\usepackage{lineno}
\usepackage{multirow}
\usepackage{xcolor}

\newcommand\um{{\textmu}m}
\newcommand\Vgain{$V_{\mathrm{gain}}$}
\newcommand\Vfd{$V_{\mathrm{fd}}$}
\newcommand\dox{$d_{\mathrm{ox}}$}

\usepackage{subfigure}
\usepackage{textcomp}
\usepackage{fouriernc}
\usepackage{here}

\journal{Nuclear Instruments and Methods A}

\begin{document}

\begin{frontmatter}



\title{Optimization of capacitive coupled Low Gain Avalanche Diode (AC-LGAD) sensors for precise time and spatial resolution}

\author[A]{Sayuka Kita}
\author[B]{Koji Nakamura}
\author[A]{Tatsuki Ueda}
\author[A]{Ikumi Goya}
\author[A]{Kazuhiko Hara}

\affiliation[A]{organization={University of Tsukuba},
            addressline={1-1-1 Tennodai}, 
            city={Tsukuba},
            postcode={305-8571}, 
            state={Ibaraki},
            country={Japan}}
\affiliation[B]{organization={High Energy Research Organization (KEK)},
            addressline={1-1 Oho}, 
            city={Tsukuba},
            postcode={305-0801}, 
            state={Ibaraki},
            country={Japan}}
\begin{abstract}
Capacitive-coupled Low-Gain Avalanche Diode (AC-LGAD) sensors are being developed for high-energy particle physics experiments as a detector which provides fast time information with fine spatial resolution. This paper describes optimizations of  AC-LGAD sensor fabrication parameters, such as doping concentrations of the gain and electrode layers as well as the AC insulator capacitance, to realize $\mathcal{O}$(10)~\um{} spacial resolution, small charge cross talk to the neighboring electrodes, detection efficiency higher than 99\% at a 10$^{-4}$ fake rate and time resolution of about 30~ps. The radiation tolerance of the sensor is presented. In addition, further application to a device capable of visible and infra-red light detection is discussed.  
\end{abstract}



\begin{keyword}
silicon tracker, LGAD, AC-LGAD, time and spatial resolution



\end{keyword}

\end{frontmatter}


\section{Introduction 
}
\label{Sec::Introduction}
Particle detectors at future lepton or hadron colliders will require to cover a very large area by a tracker with fine spatial resolution of $\mathcal{O}$(10)~\um{}. A timing capability of $\mathcal{O}$(10)~ps in addition should improve the tracking reconstruction. Precise time information is particularly inevitable in conditions where high particle density is a consequence of large multiple interactions caused by intense beam-bunch collisions. Also time can help in particle identification of charged particles, separating such between $K$/$\pi$ mesons and protons. It has been evaluated that equipping the HL-LHC ATLAS detector with 30~ps precision timing devices is significant, resulting in up to 20\% of improvement in the pileup vertices rejection~\cite{CERN-LHCC-2020-007}, which translates into a significant cost saving due to reduction in the accelerator run-time. 

The Low-Gain Avalanche Diode (LGAD) technology is for a semiconductor detector development to improve the time resolution. Presently time resolution of 30~ps ~\cite{YANG2020164379} has been achieved for minimum ionizing particles (MIPs). In realizing a spatial resolution adoptable for particle tracking, we have identified potential difficulties in improving the granularity of the electrodes~\cite{HPKDCLGADRef1,HPKDCLGADRef2}. Our first segmentation trial, a 80-\um{} pitch strip detector, showed only 20\% of the region to have a gain enough to achieve $\mathcal{O}$(10)~ps time resolution. To overcome this problem, a capacitive-coupled LGAD sensor (AC-LGAD)~\cite{HPKACLGADRef} has been developed by KEK and University of Tsukuba in collaboration with Hamamatsu Photonics K.K. (HPK). 

In this paper, optimization study of the AC-LGAD fabrication parameters, such as doping concentrations of the gain layers and the capacitance in AC electrodes, is described. The detector performance was evaluated in a laboratory measurement system and in 800~MeV electron beam. The samples were also irradiated by $^{60}$Co $\gamma$ rays and by 70~MeV protons for radiation tolerance studies. 

\section{LGAD sensors 
}
\label{Sec::LGADsensors}
The LGAD technology has now become a mature technology and 30~ps time resolution is achievable. In this section,  LGAD sensor design and improvement of granularity using AC-LGAD technology are described.

\subsection{LGAD sensor and spatial resolution 
}
\label{Sec:SpatialResolution}

\begin{figure}[btp]
    \begin{center}
    \includegraphics[width=80mm]{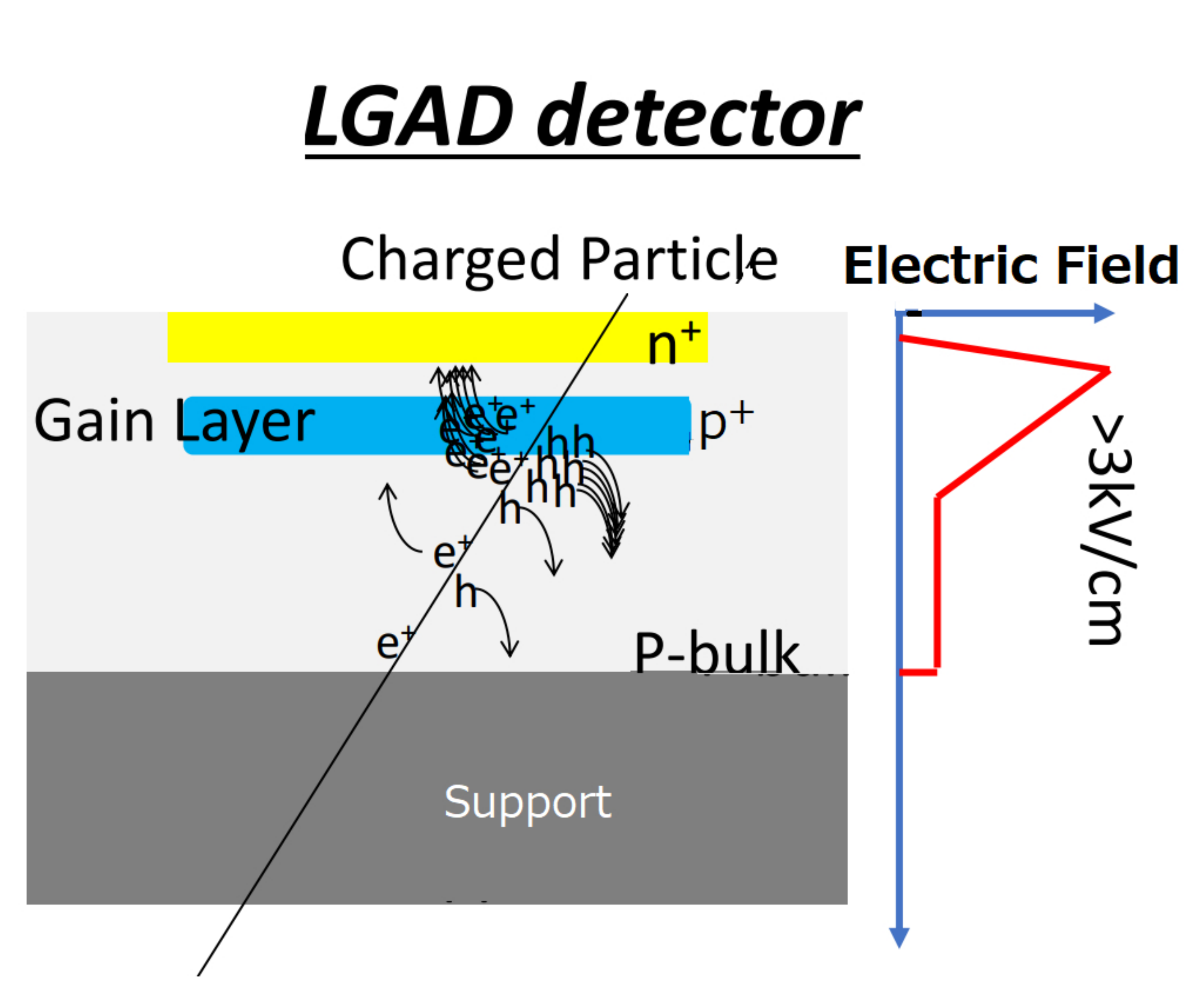}
    \caption{Operation principle of the LGAD detector}
        \label{fig:lgaden}
    \end{center}
\end{figure}

The LGAD sensor is basically an $n^+$-in-$p$ semiconductor diode, containing an additional $p^+$ layer under the $n^+$ electrodes with a larger boron doping compared to that in the $p$-bulk region. The additional layer makes an extremely higher electric field between $n^+$ and $p^+$ as illustrated in Fig.~\ref{fig:lgaden}. Such a high electric field induces avalanche multiplication, increasing the number of electron and hole pairs by about 10-20 times of that originally produced by a MIP.  Superior time resolution is achievable with LGAD as such a large signal is produced in a vicinity of the depth region.

\begin{figure}[tbp]
    \begin{center}
    \includegraphics[width=80mm]{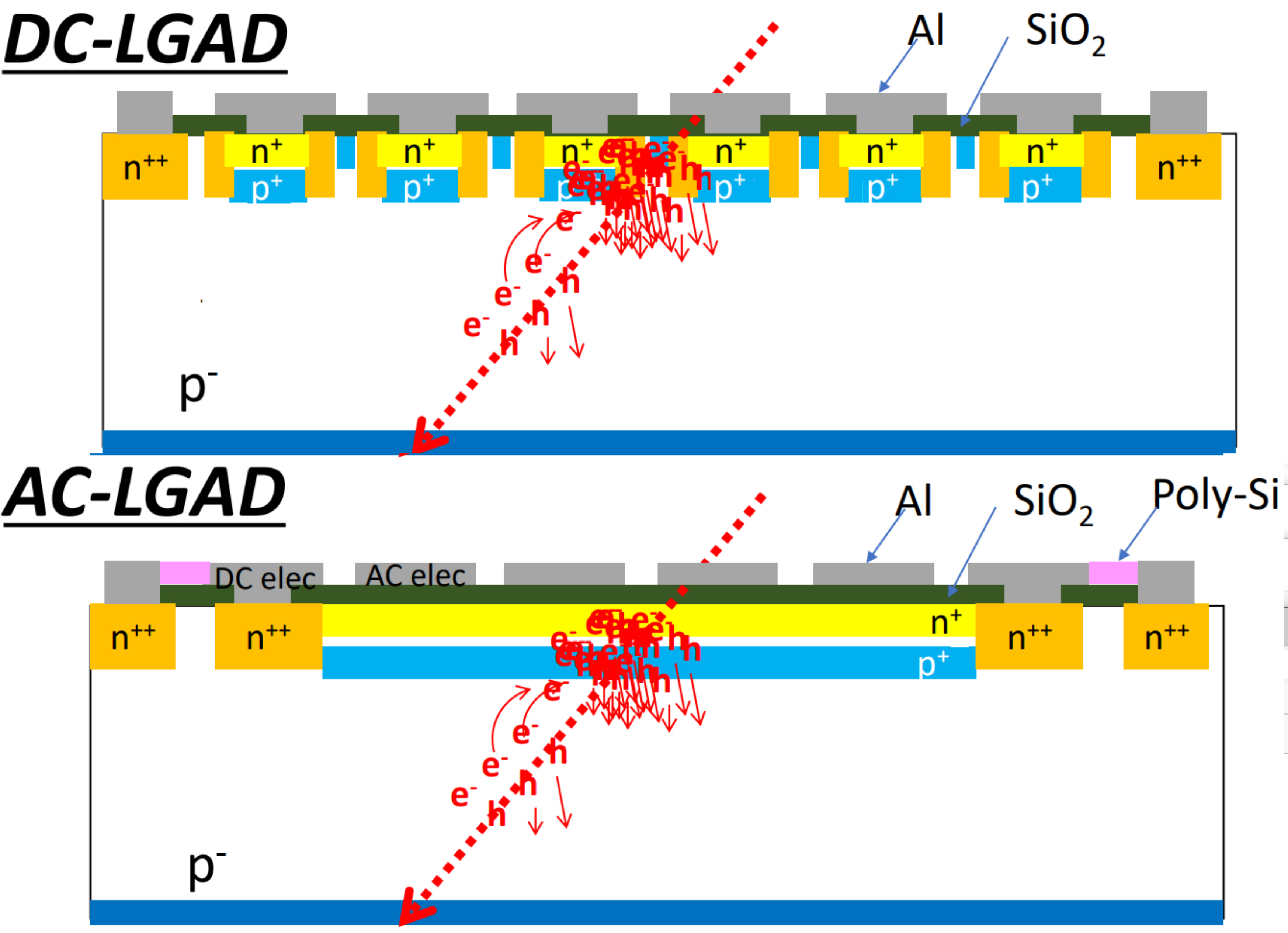}
    \caption{DC-LGAD and AC-LGAD detectors}
    \label{fig:acdc_lgad}
    \end{center}
\end{figure}

Concerning the spatial resolution improvement, the granularity of the electrodes was increased as shown in Fig.~\ref{fig:acdc_lgad}. In case where the gain layers (combination of $n^+$ and $p^+$) are formed separately for each direct-coupled electrode as shown in Fig.~\ref{fig:acdc_lgad} (top), the regions between the $n^+$ electrodes have no gain hence the fill factor is low (about 20\% for a device with 80~\um{} pitch strip type electrodes) as described in~\cite{HPKDCLGADRef2}. To overcome this feature, a capacitive-coupled LGAD (AC-LGAD) sensor, as shown in Fig.~\ref{fig:acdc_lgad} (bottom),  has been developed. 
Single uniform gain layer (combination of $n^+$ and $p^+$) is placed over the sensor active area and for the segmented readout, patterned aluminum electrodes are placed on the oxide layer and the induced signal is read out capacitively. The uniform gain layer eliminates inactive region and the avalanche multiplication occurs uniformly across the entire active area. However substantial cross-talk to the neighboring electrodes is expected due to that the charge induced in the $n^+$ layer near the avalanche point needs to escape to the ground inducing signals on the neighboring electrodes. 

\begin{figure}[tb]
\begin{center}
 \includegraphics[width=90mm]{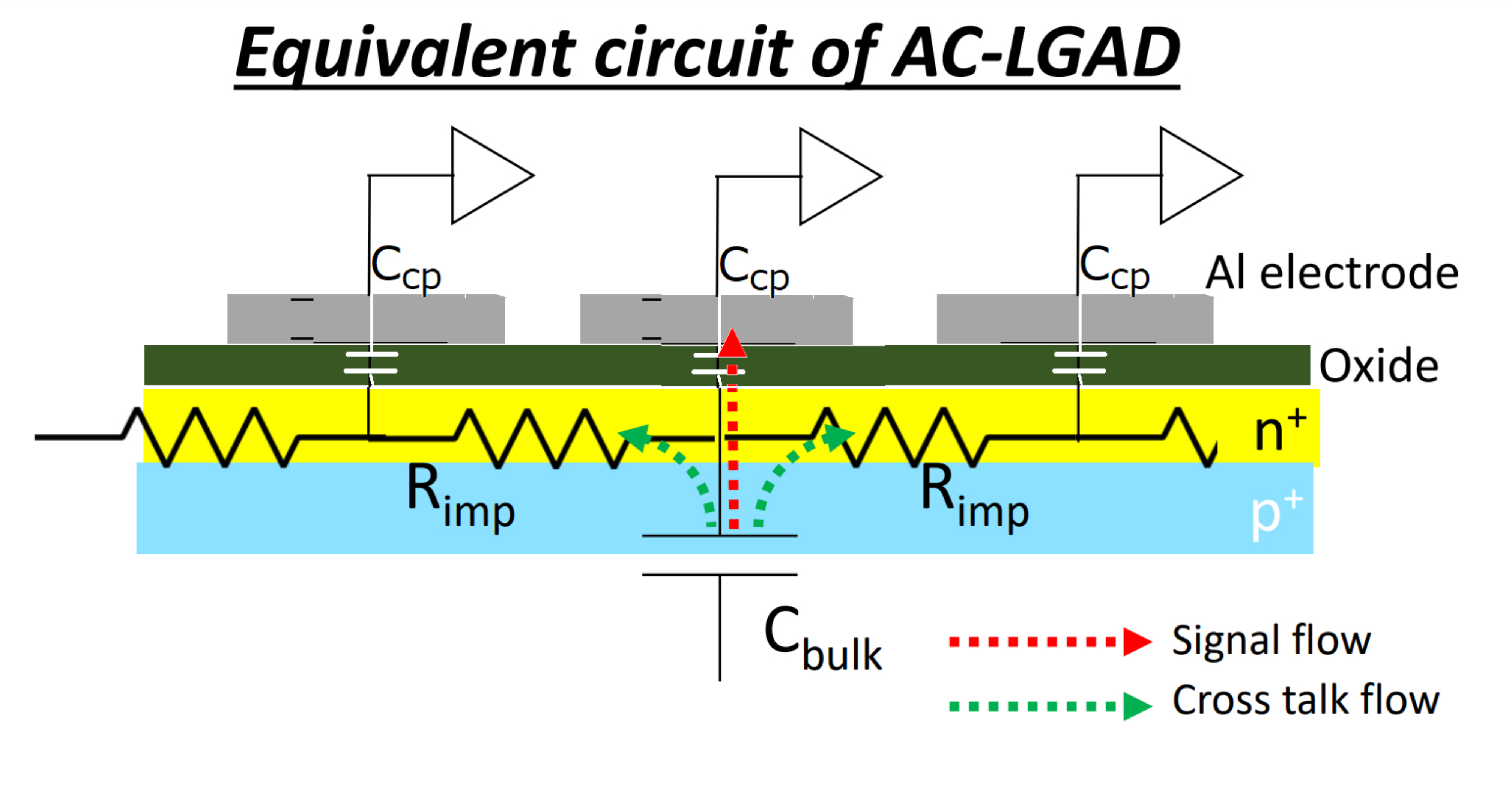}
\caption{Equivalent circuit for the signal readout in AC-LGAD sensor. Signal flow and cross-talk flow are shown by red and green dotted arrows, respectively.}
\label{fig:ccprimp}
\end{center}
\end{figure}

The signal charge produced by the avalanche ($Q_{\mathrm{0}}$) flows in two paths, one is read out via capacitive coupling ($C_{\mathrm{cp}}$), and the other is cross-talk through the $n^+$ implant that has an effective resistance ($R_{\mathrm{imp}}$), as illustrated in Fig.~\ref{fig:ccprimp}. The readout charge ($Q$) is determined by the two components, $Z_{C_{\mathrm{cp}}}$ and $Z_{R_{\mathrm{imp}}}$, which are the impedance of $C_{\mathrm{cp}}$ and $R_{\mathrm{imp}}$, respectively, as in the following equations:

\begin{equation}
Q = \frac{Z_{R_{\mathrm{imp}}}}{Z_{C_{\mathrm{cp}}}+Z_{R_{\mathrm{imp}}}}\times Q_0.
\label{e1}
\end{equation}
\begin{equation}
Q = \frac{R_{\mathrm{imp}}}{\sqrt{1/(2\pi f C_{\mathrm{cp}})^2+R_{\mathrm{imp}}^2}}\times Q_0,
\label{e1}
\end{equation}
where $f$ is effective frequency of the pulse signal. Larger the $C_{\mathrm{cp}}$ and $R_{\mathrm{imp}}$,  higher the pulse height signal and lower the cross-talk. This model indicates that the signal and cross-talk sizes are controllable by adjusting the doping concentration of $n^+$ layer: as reducing the doping concentration results in higher resistance, larger signal and lower cross-talk are expected.
The thickness of oxide layer between the aluminum electrode and $n^+$ layer changes the $C_{\mathrm{cp}}$: reducing the oxide thickness (\dox{}) results in larger signal and lower cross-talk. 
The doping concentration of $p^+$ layer controls the gain. The operation    voltage, $e.g.$, \Vgain{} as defined below, must be higher than the full depletion voltage (\Vfd) while higher initial \Vgain{} results in worse radiation tolerance as described in Section~\ref{sec:radiationtolerance}.

In high particle density environment like at high instantaneous luminosity hadron colliders, lower cross-talk is preferred not to increase the hit occupancy of the tracking detectors. On the other hand, in experiments in relatively lower particle flux environment like at the electron--ion collider (EIC) \cite{eic}, the cross-talk information is actively used to determine the particle position from the signal sharing between two neighboring electrodes where the relationship of the signal amplitude ratio and particle position is parameterized. For this purpose, typically 30-50\% of charge cross-talk is optimum for the spatial resolution with relatively large electrode size e.g. 500~\um{}~\cite{Heller_2022}.
In the following, optimization has been performed aiming at achieving low cross-talk for future high luminosity hadron colliders.

\subsection{AC-LGAD prototype production at HPK 
}
\label{Sec::ACLGADPrototype}
According to the charge flow model, the sensor performance is determined by the size of the electrode, the doping concentrations and oxide thickness \dox{}. Three sensor types with different electrode sizes: pad, strip and pixel, and eleven types of different doping concentrations and \dox{}'s have been designed and fabricated by HPK. 

Table~\ref{tab:sensorgeotype} summarizes the physical parameters of the produced sensor types: the overall dimension, electrode pitch, electrode size and the number of electrodes. The active $p$-bulk thickness is 50~$\mu$m for all the samples. To optimize the electrode size, the aluminum electrode size is varied for the strip and pixel types with fixed electrode pitches. 
\begin{table}[tb]
\begin{center}
\caption{The physical parameters of the produced sensor types; Pad, Strip and Pixel. The overall size, electrode pitch, electrode width, and the numbers of electrodes in column and row directions are summarized.}
\label{tab:sensorgeotype}
\begin{tabular}{l|ccc}
\hline
\hline
Sensor type & Pad & Strip & Pixel \\
\hline
overall (mm) 
&3 $\times$ 3  & 12 $\times$ 6.7  & 2.1 $\times$ 1.8  \\
\hline
electrode
 &  500$\times$500 & 80 & 50$\times$50 \\
pitch (\um{})  & &  & \\
\hline
electrode &   450 &30, 35, 40, 45 & 30$\times$30, 34$\times$34, \\
 size (\um{})  & $\times$450 & $\times$9880 & 38$\times$38, 42$\times$42 \\
\hline
column$\times$row  & 2$\times$2  &  48 &  14$\times$14  \\

\hline
\hline
\end{tabular}
\end{center}
\end{table}

The three parameters, \dox{}, $n^+$  and  $p^+$ doping concentrations, were subjected to change as summarized in Table~\ref{tab:sensortype}. In total eleven variations in the doping concentrations and oxide thickness were examined. The B samples were produced with three different $p^+$ doping concentrations to evaluate the \Vgain{} dependency on the $p^+$ doping concentration. The $p^+$ doping concentration of D and E samples was tuned such that \Vgain{} is roughly at 170~V. The types with "a" and "b" refer two types of \dox{} values with "b" having 1.5 times higher capacitance than "a". The others are all with "a".

\begin{table}[tb]
\begin{center}
\caption{Summary of eleven prototype variations in $p^+$ and $n^+$ implant concentrations and oxide thicknesses: five types of $n^+$ resistivity (A to E) and three types of $p^{+}$ doping concentration for A to C samples (1: low, 2: middle, 3:high). For D and E types, $p^{+}$ doping concentration is tuned so that gain voltage is roughly at 170~V. For the samples with "a" and "b", two coupling capacitances are produced; capacitance of"b" type is 1.5 times larger than "a". Others are with the default capacitance of "a". }
\label{tab:sensortype}
\begin{tabular}{c|ccccc}
\hline
\hline
 & \multicolumn{5}{c}{\textbf{n$^+$ resistivity} [$\Omega/\square$] } \\
  p$^{+}$ doping conc.& 1600  & 800 &400 & 132 &  40  \\
\hline
   high   &       &       &       & B-3 & A-3 \\
   middle &       &       & C-2/b & B-2 & A-2 \\
   (170V) &       &D-a/b  &       &     &  \\
   (170V) & E-b   &       &       &     &  \\
   low    &       &       &  C-1  & B-1 &  \\
\hline
\hline
\end{tabular}
\end{center}
\end{table}

\section{Leakage current and gain voltage 
}
\label{Sec::LeakageCurrent}


The leakage current vs. bias voltage, $IV$, dependence was measured at first to extract the operation voltage. Negative bias voltage was supplied to the backside of the sensor with grounding the DC ring which is electrically connected to the $n^+$ gain layer via poly-Si bias resistor of 1-2~M$\Omega$. 
The $IV$ curve was measured at a 2~V voltage step at 20$^\circ$C.

\begin{figure}[htbp]
    \begin{center}
\includegraphics[width=65mm]{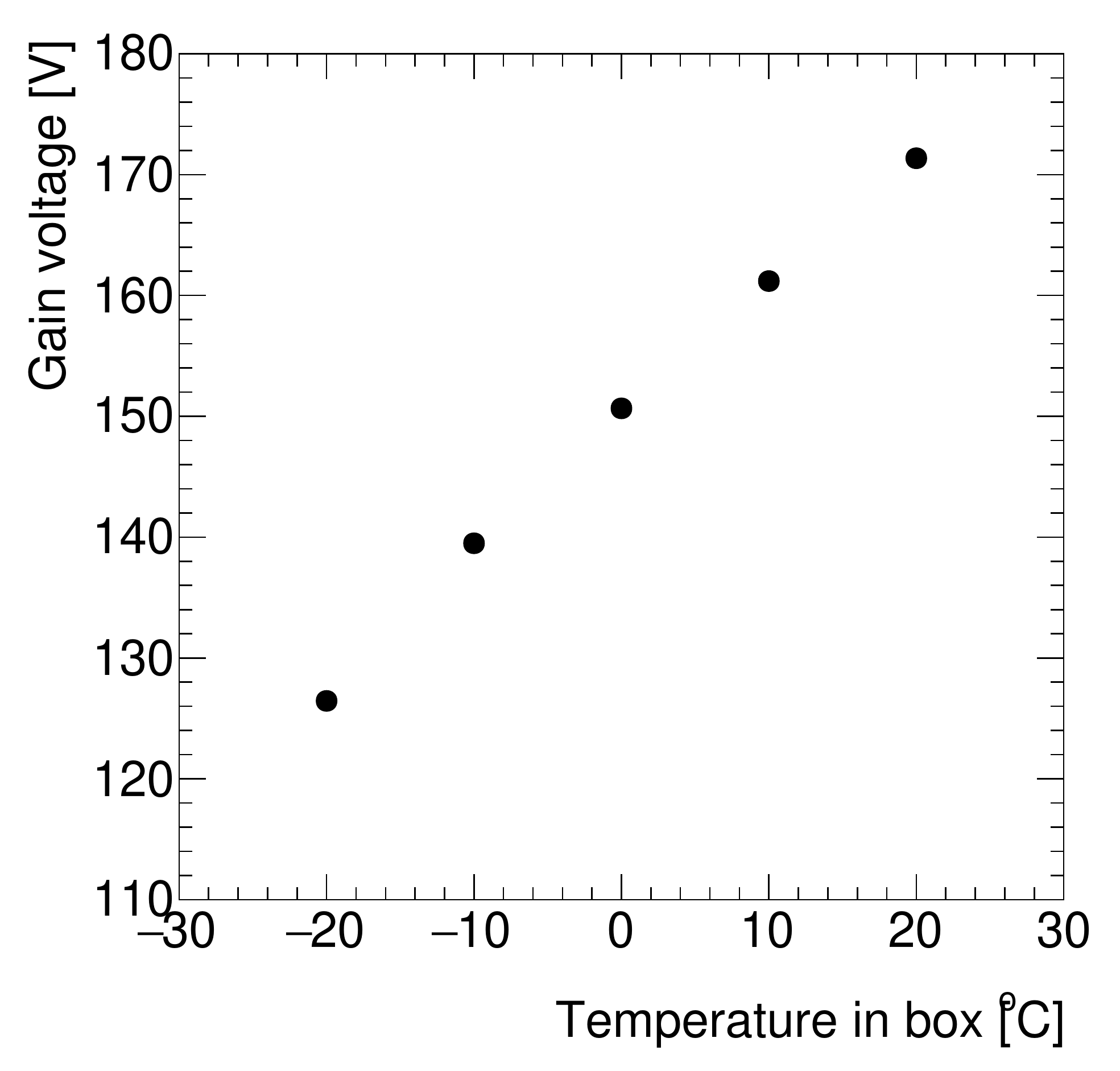}

   \caption{Temperature dependence of the gain voltage obtained by E-b type sensor.}
   \label{fig:IVthemo} 
    \end{center}
\end{figure}

\begin{figure}[h]
    \begin{center}
    \includegraphics[width=70mm]{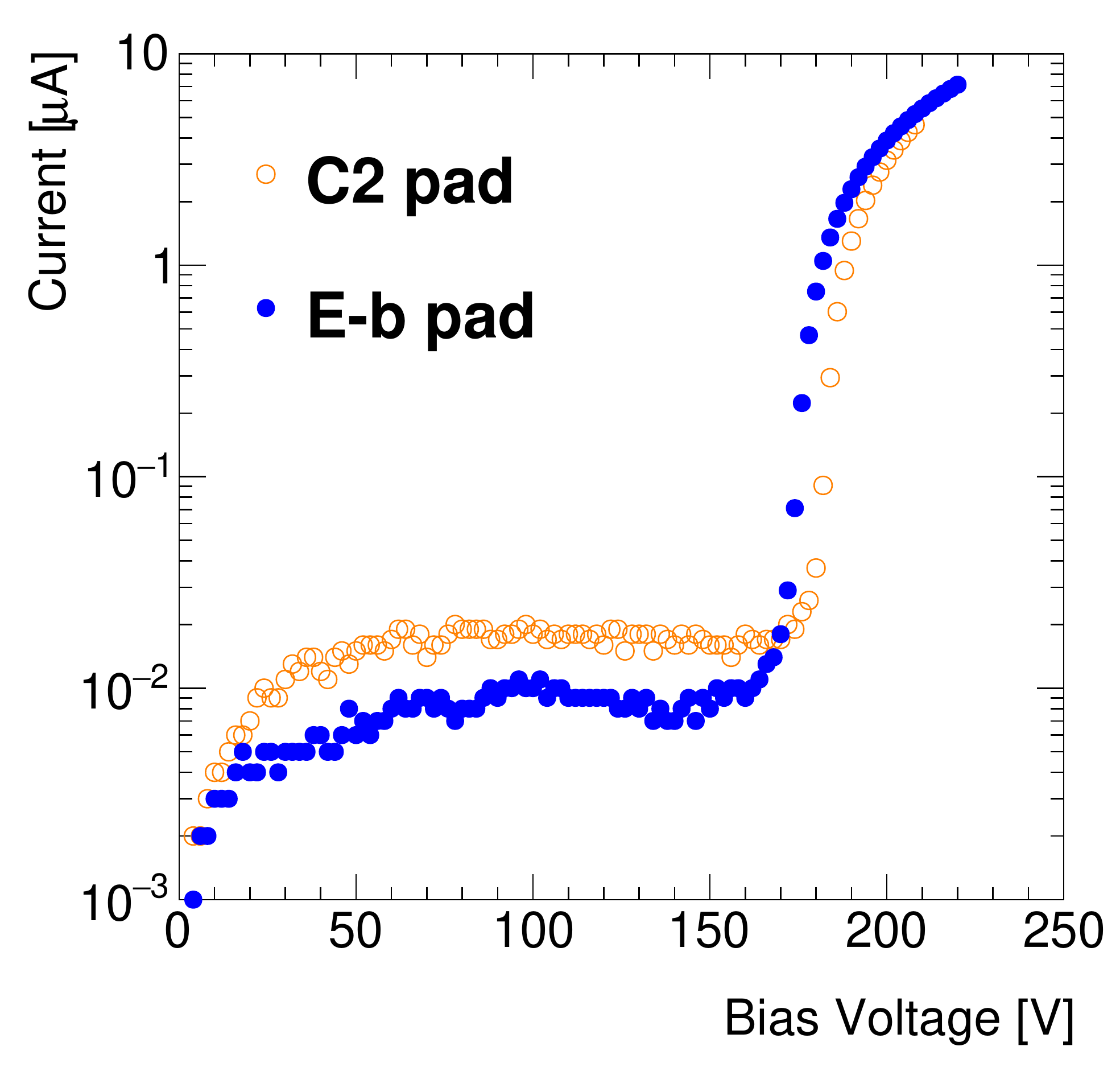}
    \caption{ $IV$ curves of C-2 and E-b pad type samples.}
    \label{fig:IVC2Eb}
    \end{center}
\end{figure}

 The pad type sensors of eleven variations were used to compare \Vgain{} which is defined as the bias voltage at which is evaluated by crossing point of two different fitted linear functions for lower and higher than the point of leakage current rapidly increase regions.
 Table~\ref{tab:vbdcomp} summarizes the gain voltages for the A to C samples at room temperature. The sensor with higher $p^+$ doping concentration gives a lower \Vgain{} as expected. The $n^+$ doping concentration dependence is such that the sample with larger $n^+$ resistivity shows a lower \Vgain{}. The temperature dependence of \Vgain{} is observed as shown in Fig.~\ref{fig:IVthemo}. The \Vgain value increases with temperature with a slope ($\Delta$\Vgain{}/${\Delta}T$) $\sim$ 1.1~V/$^{\circ}$C. Fig.~\ref{fig:IVC2Eb} shows the $IV$ curves for the E-b type and C-2 type measured at 20$^{\circ}$C temperature. 
 \Vgain{} is 174~V for E-b type and  182~V for C-2 type .

\begin{table}[tbh]
\begin{center}
\caption{Gain voltage \Vgain{} of A to C pad types measured at 25$^{\circ}$C. }
\label{tab:vbdcomp}
\begin{tabular}{c|ccc}
\hline
\hline
& \multicolumn{3}{c}{\textbf{n$^+$ resistivity} } \\
$p^+$ doping conc.&C & B &  A  \\
\hline
                high&   & 113 & 236\\
                middle&   185 & 234 & 264 \\
    low& 283 & 318 & \\
\hline
\hline
\end{tabular}
\end{center}
\end{table}

\section{Characterization using $\beta$ rays 
}
The sensor performance of detecting MIP signals was evaluated using $\beta$-rays off strontium 90 ($^{90}$Sr). The signals are to be well separated from the noise and the cross-talk needs to be small enough such that the traverse position of $\beta$-ray is well reconstructed. In this section, the signal size and cross-talk size are discussed. 


\subsection{Measurement setup 
}\label{Measurementsetup}
\begin{figure}[htbp]
    \begin{center}
\includegraphics[width=80mm]{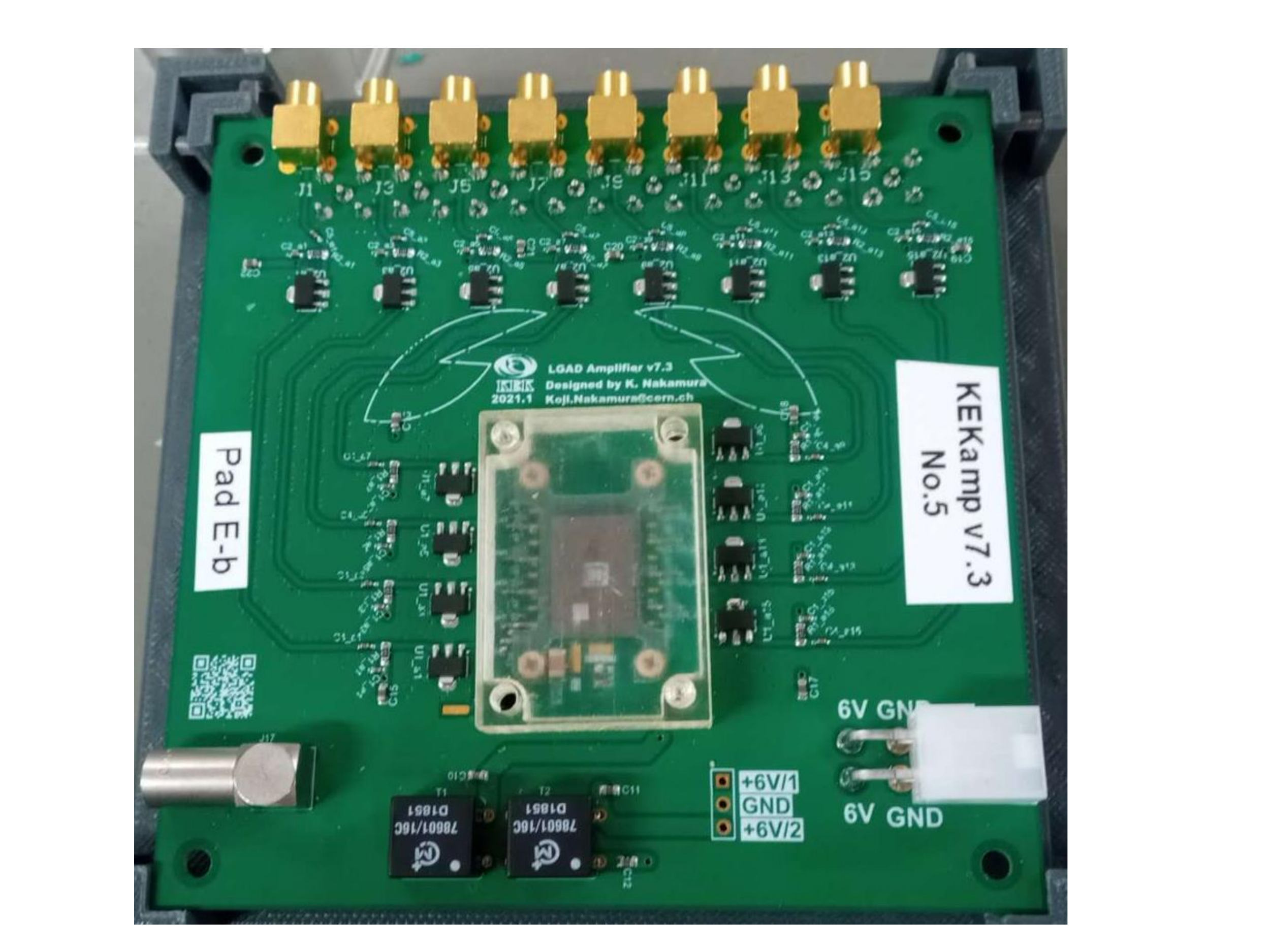}
 
    \caption{KEK 16-ch amplifier board.}\label{fig:kaniam}
    \end{center}
\end{figure}

The measurement of signal from the sensor requires a fast amplifier and a high-speed digitizer. A discrete amplifire readout board comprising 16-channel amplifiers is developed, each with two-stage charge amplifier integration circuits (IC), as shown in Fig.~\ref{fig:kaniam}. One sensor sample was mounted on the amplifier board via electrically conductive tape in order to supply bias voltage to the backside of sensor. The bias voltage was supplied by Keithley 6517 or 2410. The AC-LGAD readout electrodes are wire-bonded to the amplifier inputs. 

The amplified signals are extracted and digitized by a desktop 16-channel digitizer CAEN DT5742 (5~GHz sampling rate with 10-bit time memory cells, each with 12-bit ADC for 1~V full scale) or an 8-channel oscilloscope LeCroy WaveRunner 8208HD (10~GHz sampling with 12-bit ADCs). 
While placing a $^{90}$Sr source on top of the sensor, the pulse signals are digitized on receiving a trigger from a system consisting of a scintillator and two Multi-Pixel Photon Counters (MPPC, S13360-1350CS \cite{HPK_MPPC}) set under the amplifier board. The whole setup is placed in a climate chamber to keep the sensor temperature at 20$^\circ$C.

\subsection{Pulse height measurement 
}
\label{sec:phmeasure}
\begin{figure}[htbp]
    \begin{center}
\includegraphics[width=80mm]{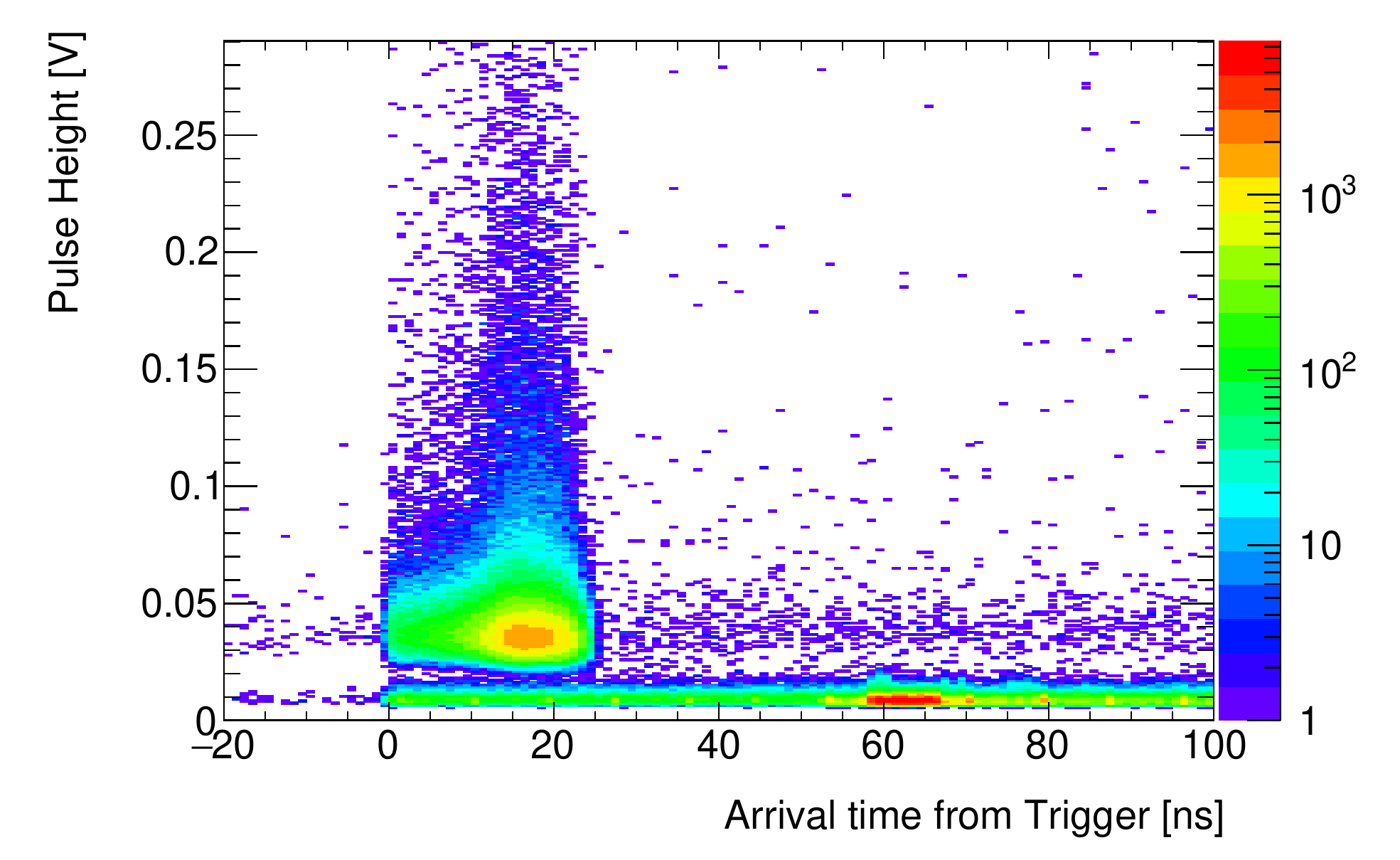}

  \caption{Pulse height and arrival time distribution for Strip E-b type.} \label{fig:phvstime}
\end{center}
\end{figure}

\begin{figure}[h]
    \begin{center}   
    \includegraphics[width=80mm]{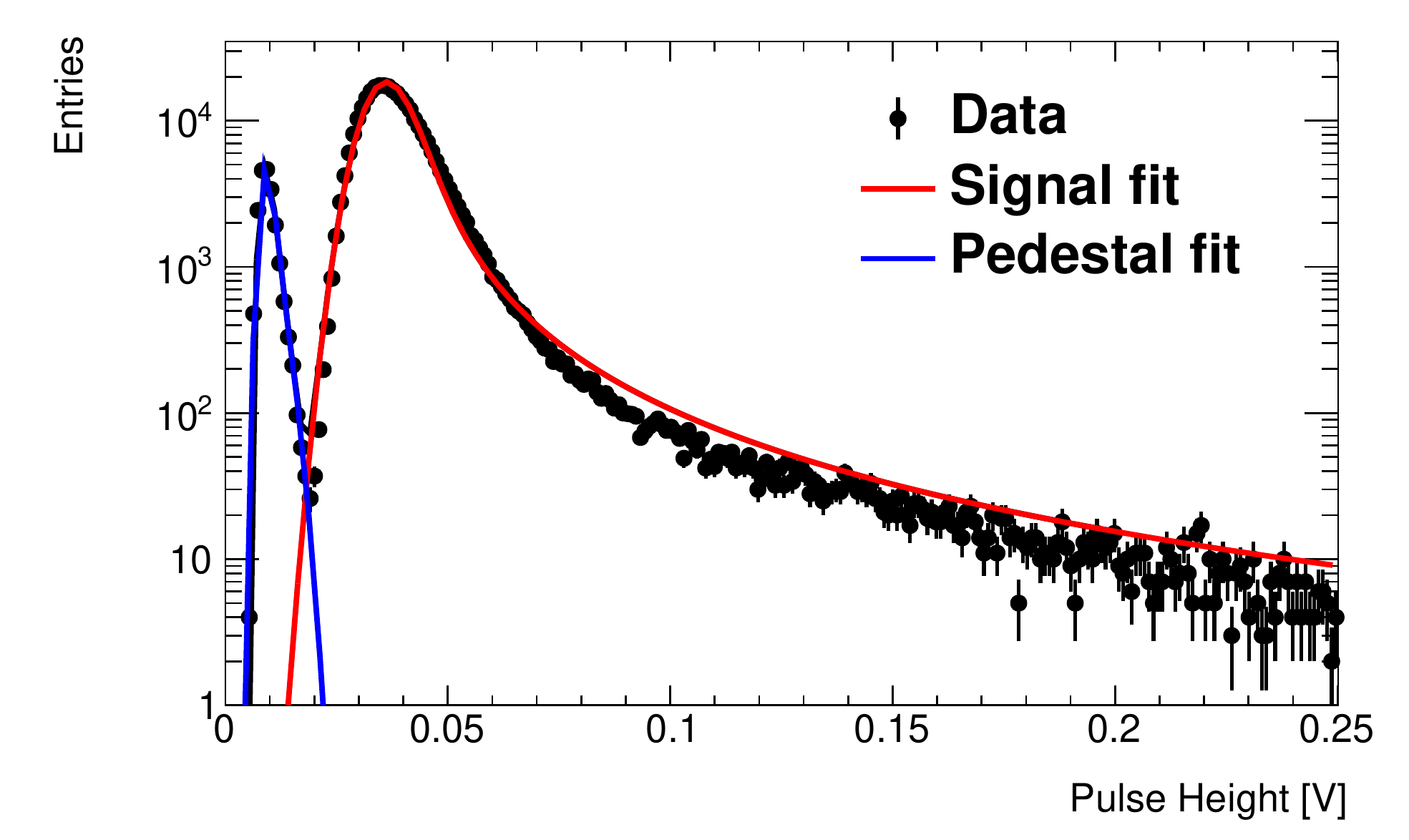}
    \caption{Distributions of on-time and off-time pulse heights for Strip E-b type (strip pitch = 80 {\textmu}m).}\label{fig:setuppicEbSig}
    \end{center}
\end{figure}

The amplified sensor signal is a negative pulse with small positive overshoot (example is in \cite{HPKACLGADRef} Fig. 2). The pulse height is defined as the largest negative voltage among the sampled ADC values with respect to the base line. The arrival time of the signal is defined as the time at 50\% of the pulse height, as a constant fraction threshold. Fig.~\ref{fig:phvstime} shows the pulse height of E-b strip type sensor at 176~V plotted against the arrival time relative to the trigger. On-time events are defined as those in the region with a time difference less than 30~ns relative to the trigger. The events in the region with arrival time difference more than 30~ns and less than 50~ns are defined as off-time events. The off-time events are used to parameterize the pedestal distributions which are characterized by the intrinsic noise due to the sensor leakage current, the amplifier and digitizer noises. Fig.~\ref{fig:setuppicEbSig} shows the on-time pulse height distribution while the fitted curves 
correspond to the pedestal and signal components.  The off-time pulse height distribution for the pedestal is fitted with an asymmetric Gaussian function. For the signal distributions, a Gaussian convoluted Landau function was used. The most probable values  (MPVs) of the signal function is defined as the signal size in the following.

\subsection{Signal size 
}
The signal pulse height distributions for E-b pad and strip type sensors are compared. 
As the results, pulse height MPVs of pad sensor and strip sensor are 117.8 $\pm$ 1.4 mV and 34.46 $\pm$ 0.02 mV respectively. 

\begin{figure}[htbp]
    \begin{center}   
     \includegraphics[width=70mm]{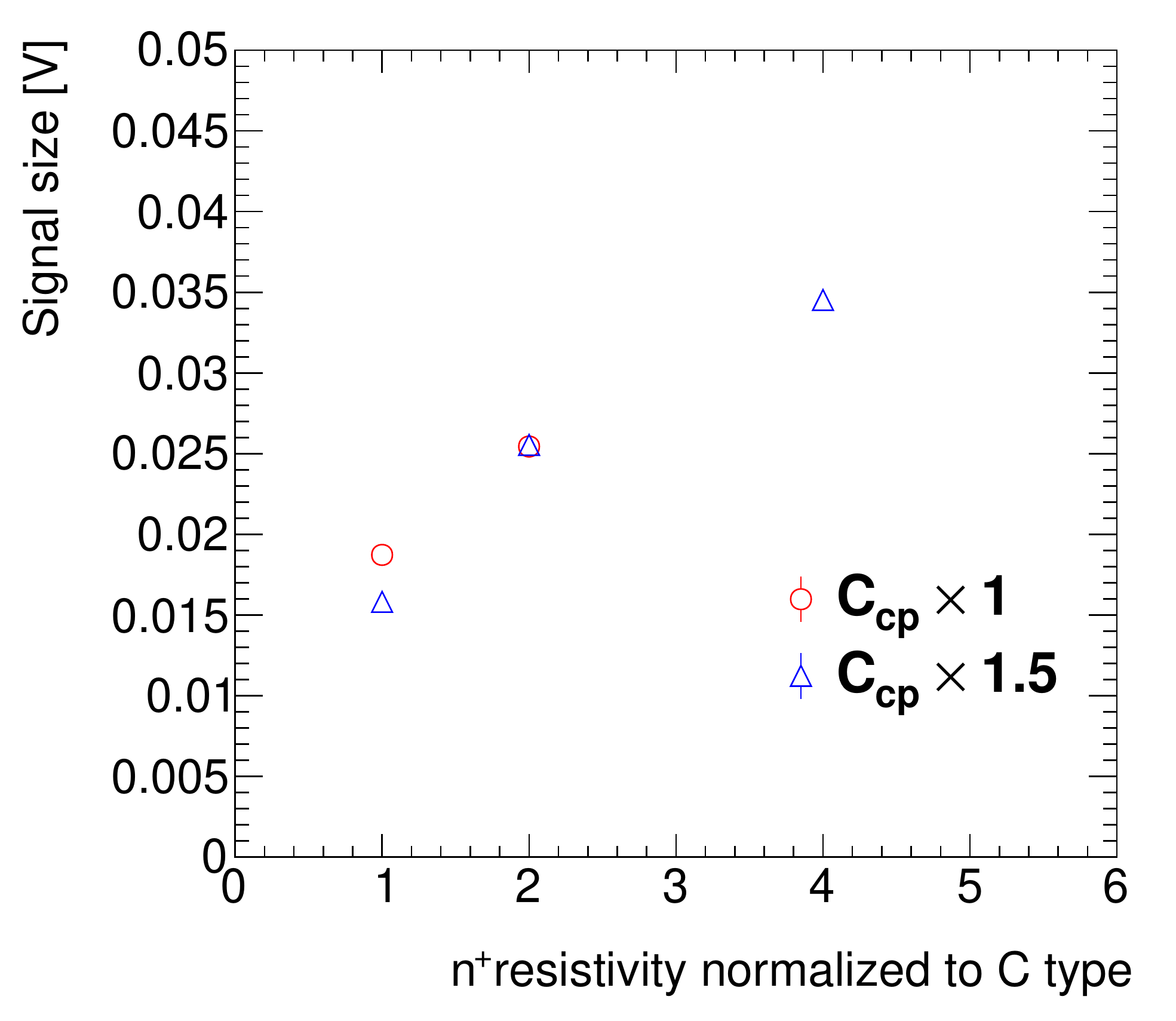}
    \caption{Signal size as a function of the $n^+$ implant resistivity normalized to that of C type.}
    \label{fig:sigsize}
    \end{center}
\end{figure}

To evaluate the $n^+$ resistivity dependence of the signal size, strip sensors C-2, C-2b, D-a, D-b and E-b types were compared. Fig.~\ref{fig:sigsize} shows the signal size as a function of the $n^+$ implant resistivity normalized to that of C-2 type. Higher resistivity results in higher signal size as explainable by the signal readout model described in Section~\ref{Sec:SpatialResolution}. 
The signal separation from pedestal is evaluated by the highest resistivity sample, E-b type. The efficiency value 99.98\% was obtained at the threshold which makes noise rate 10$^{-4}$. 


\subsection{Cross-talk 
}
\begin{figure}[h]
    \begin{center}                        
     \includegraphics[width=80mm]{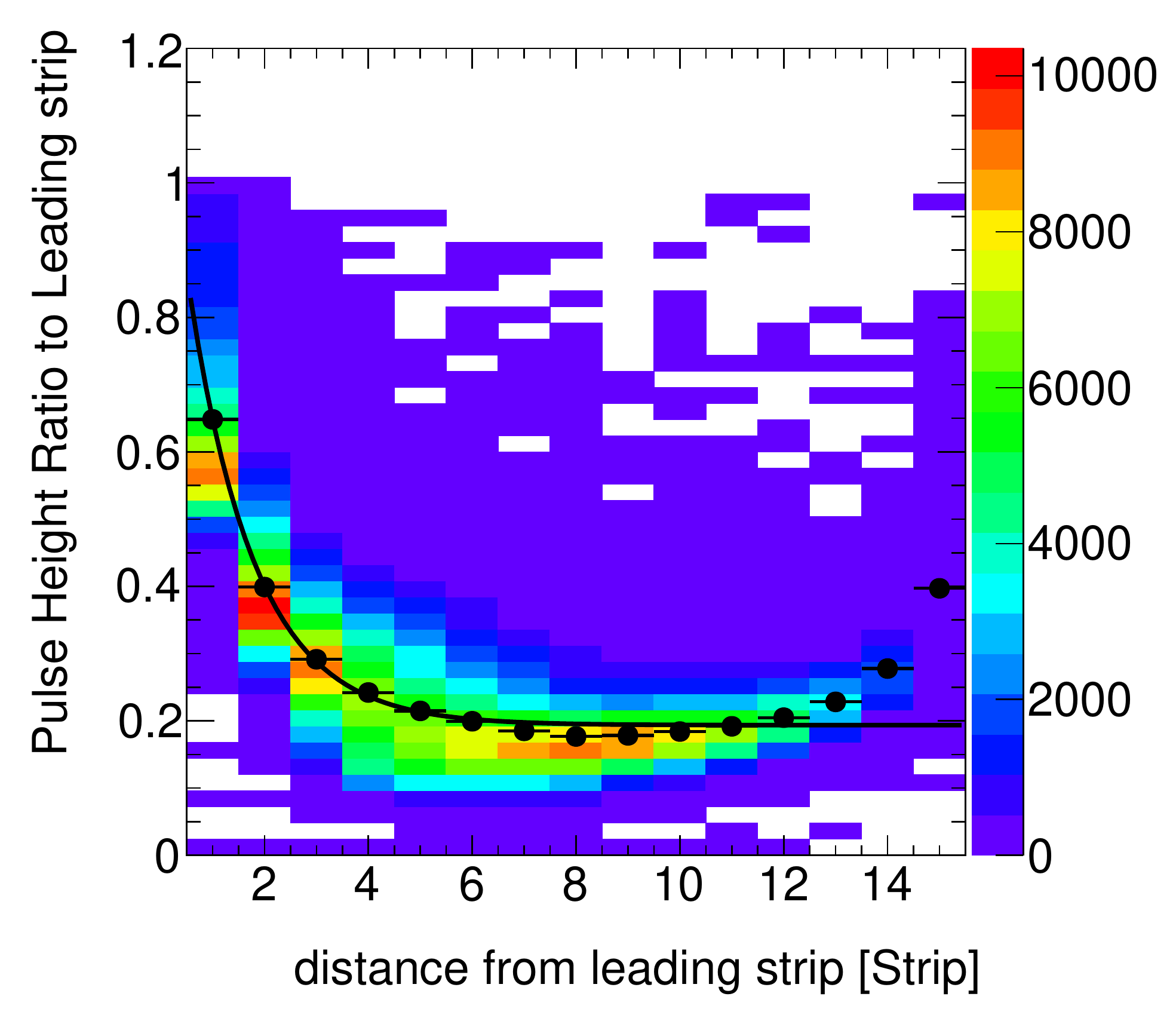}
    \caption{Pulse height ratio to the leading channel pulse height as function of distance from the leading. Strip E-b type (strip pitch = 80~$\mu$m).}
    \label{fig:xtalkdist}
    \end{center}
\end{figure}

One of the critical issues in designing an AC-LGAD sensor is to suppress the cross-talk to neighboring electrodes to a manageable level, as it is expected to occur via $n^+$ implant which is uniformly placed. 
The channel with the highest pulse height in an event was defined as the leading channel and the pulse height ratios were derived for the other channels as the pulse heights relative to that in the leading channel. The magnitude of the cross-talk was measured for the strip type sensors. The pulse height ratio as a function of distance from the leading strip channel is shown in Fig.~\ref{fig:xtalkdist} for E-b type. 

\begin{figure}[h]
    \begin{center}                        
    \includegraphics[width=70mm]{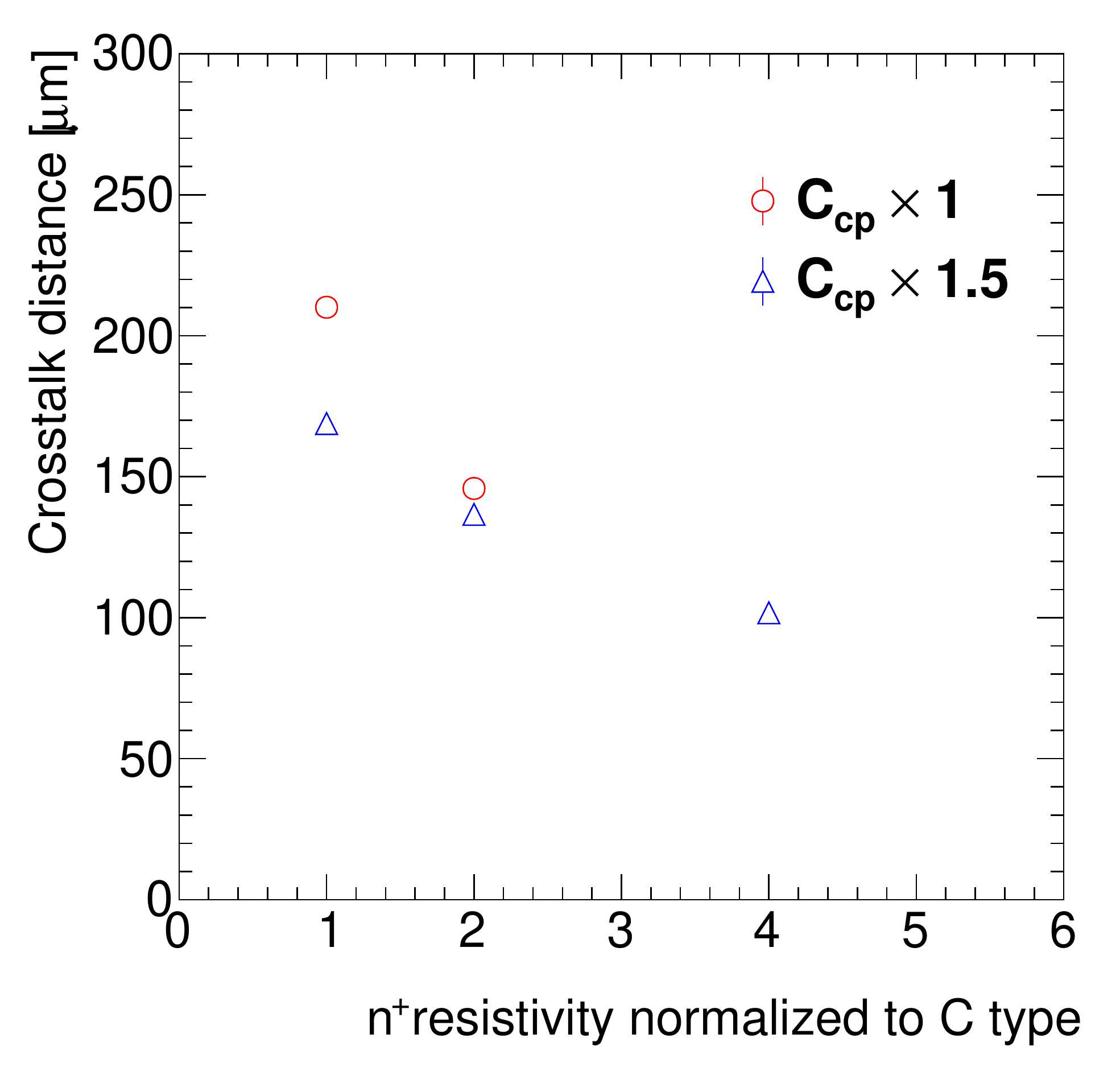}
    \caption{Cross-talk distance as function of the $n^+$ resistivity normalized to that of C type.}
    \label{fig:xtalkdistcomp}
    \end{center}
\end{figure}

The mean of the ratios is calculated and its dependence on the distance is fitted to a function of exponential+constant: fitted function is overlaid in Fig.~\ref{fig:xtalkdist}. We define the cross-talk distance as the distance constant of the exponential.
The cross-talk distances are plotted in Fig.~\ref{fig:xtalkdistcomp} as a function of the $n^+$ implant resistivity. As expected, the cross-talk distance becomes shorter with the resistivity.


\subsection{Time resolution  
}
The time resolution of the AC-LGAD detectors has been evaluated using high-energy proton beam and the result for C-2 pad sensor is reported in Fig.~12 (right) in~\cite{Heller_2022}. No significant position dependence including in the gap of the electrodes is observed. The time resolution is 30$\pm$1~ps derived using a 20\% constant fraction threshold algorithm.

As the accessibility to testbeam is limited,  a time resolution measurement setup using $^{90}$Sr source has been developed. The setup is effective to shorten the detector development cycle. 

\begin{figure}[htbp]
    \begin{center}   
    \includegraphics[width=70mm]{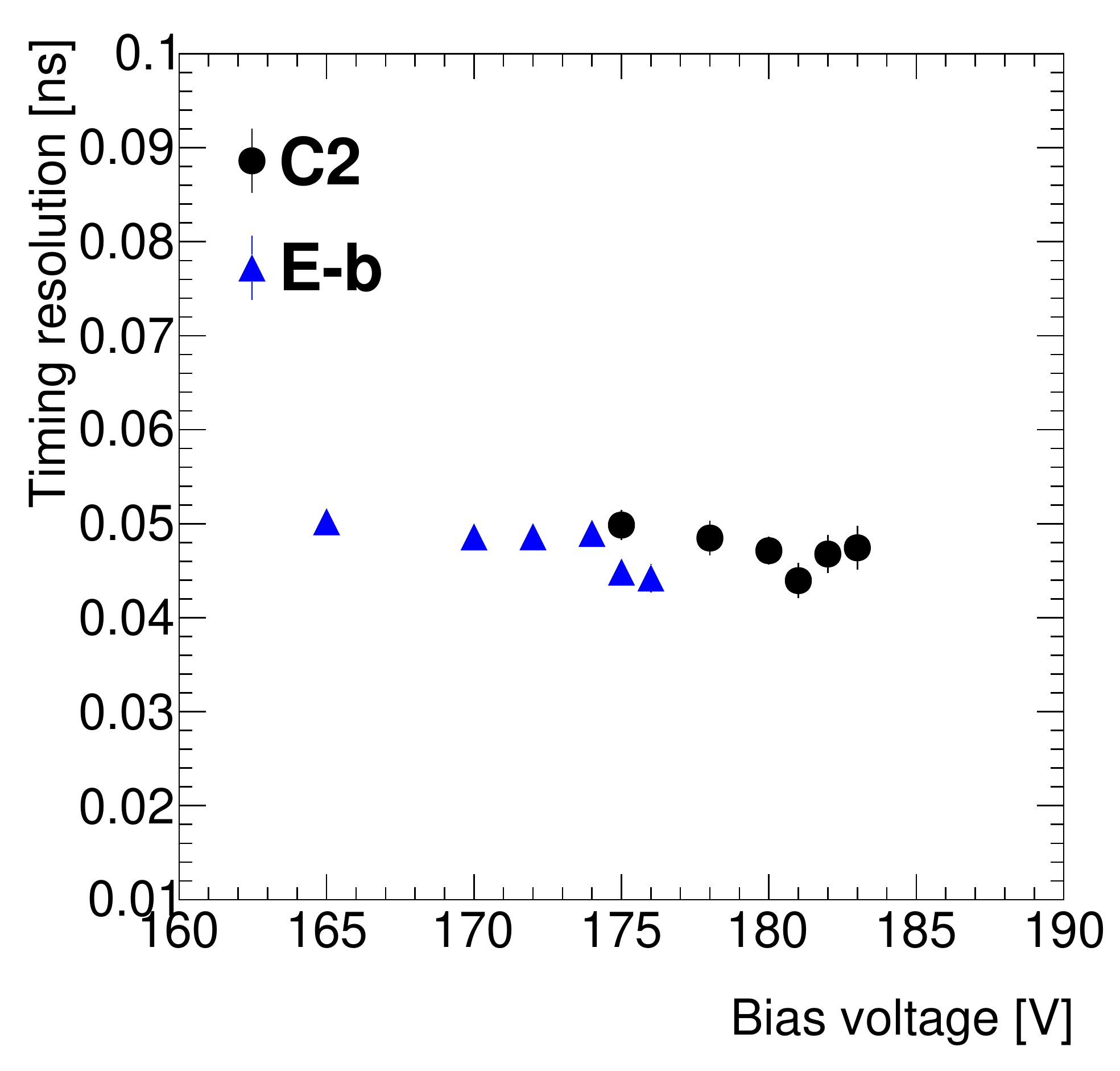}
    \caption{Time resolution as function of bias voltage of pad-type C2 and E-b sensors.}
    \label{fig:timereso}
    \end{center}
\end{figure}

Two amplifier boards are stacked with a 1~cm gap between. One board is placed on a manually adjustable stage so to align two sensors precisely. 
The distribution of time differences recorded by the two sensors of identical type with applying a same bias voltage is fitted to a gaussian. The standard deviation over $\sqrt{2}$ of the gaussian corresponds to the time resolution at that bias. Fig.~\ref{fig:timereso} plots the time resolution as a function of the bias voltage measured for two types of pad detector, C-2 and E-b. The time resolutions of E-b and C-2 are 44.2$\pm$1.5~ps at 176~V and 43.9$\pm$1.9~ps at 181~V, respectively. Although the results are worse than the testbeam results, 
the time resolution of sensors of various parameters can be compared in a laboratory test setup.

\section{Characterization using ELPH electron beam 
}
The efficiency and spatial resolution were evaluated using an 800~MeV electron beam available at Research Center for Electron Photon Science (ELPH), Tohoku University  \cite{ELPH}.

\subsection{Setup 
}

\begin{figure}[b]
    \begin{center}
        \includegraphics[width=80mm]{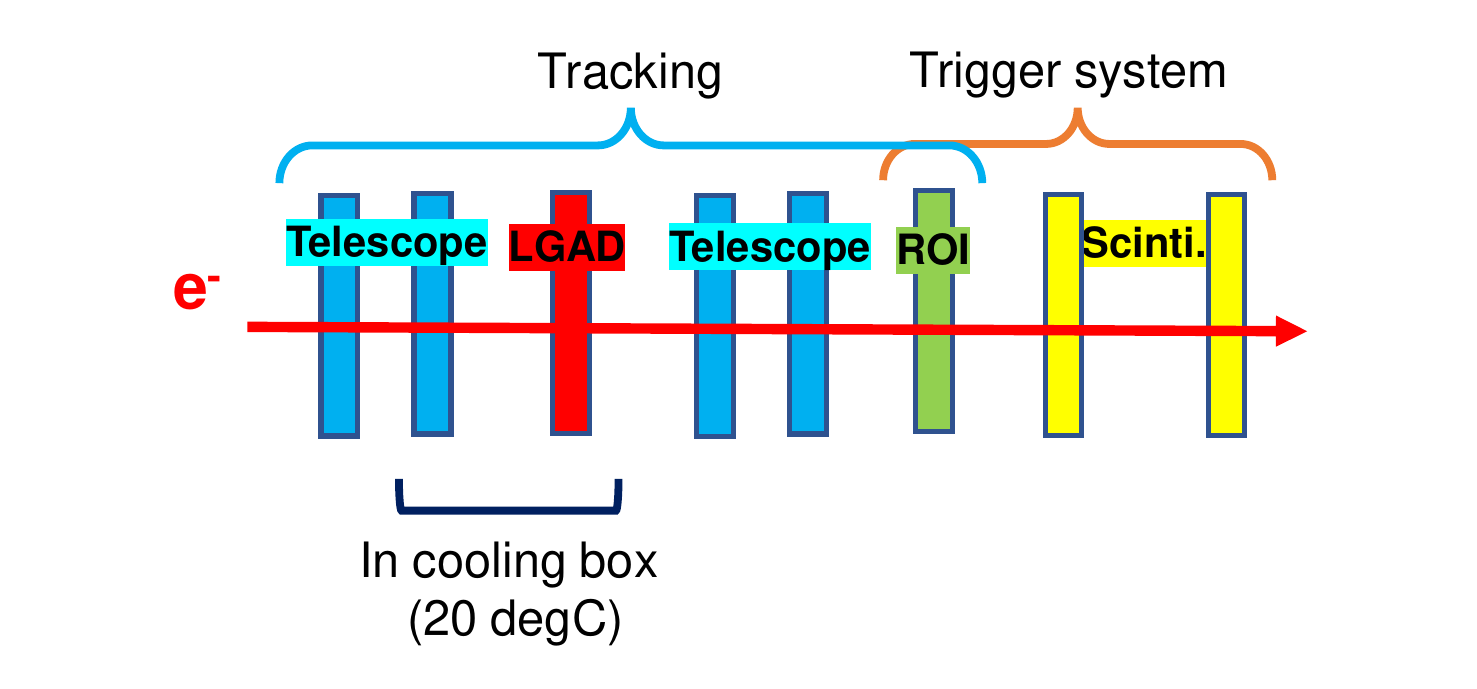}
        \caption{ELPH testbeam setup.  The testing sensors are placed inside the thermal box.} 
        \label{fig:tbsetuppic}
    \end{center}
\end{figure}

The ELPH provided an 800 MeV electron beam at 200$\sim$400~Hz beam rate available to the setup shown in Fig.~\ref{fig:tbsetuppic}. 
Two scintillator planes (20~mm~$\times$~30~mm active area) read out with multi-pixel photon counters (MPPCs)
were placed at the downstream end of the setup. Five telescope planes were used for beam tracking. The ATLAS FE-I4 based modules~\cite{IBL_collaboration_2012} were comprised with four planes with pixel size of 25$\times$500~{\textmu}m$^2$ and one with pixel size of 50$\times$250~{\textmu}m$^2$. The four planes with 25$\times$500~{\textmu}m$^2$ pixel size sensors were rotated by 90$^\circ$ for the second and fourth in order to achieve fine tracking reconstruction in both directions. 
The 50$\times$250~{\textmu}m$^2$ pixel size plane generated a trigger when the particle traversed a certain area of the plane defining the region of interest (ROI). This ROI trigger was helpful for testing small area of the LGAD under test.  
The E-b type LGAD sensor was placed between the second and third telescope planes. The LGAD sensor and the second telescope plane were set in a thermal box to keep the temperature at 20$^\circ$C using a three-stage Peltier based cooling system. 

Xilinx Kintex-7 FPGA KC705 was used as the trigger logic unit (TLU) to construct the trigger. Logical AND of ROI and two scintillator signals was the trigger with succeeding triggers vetoed by the sum of busy signals from the data acquisition systems including of the FE-I4s and the digitizer.

Events with hits in all telescope planes were retained. The track was reconstructed by fitting the hit points to a straight line and good tracks were defined as those with reduced chi-squares less than 18 taking multiple scattering effect into account.

\subsection{Signal efficiency 
}
\begin{figure}[h]
    \begin{center}
\includegraphics[width=70mm]{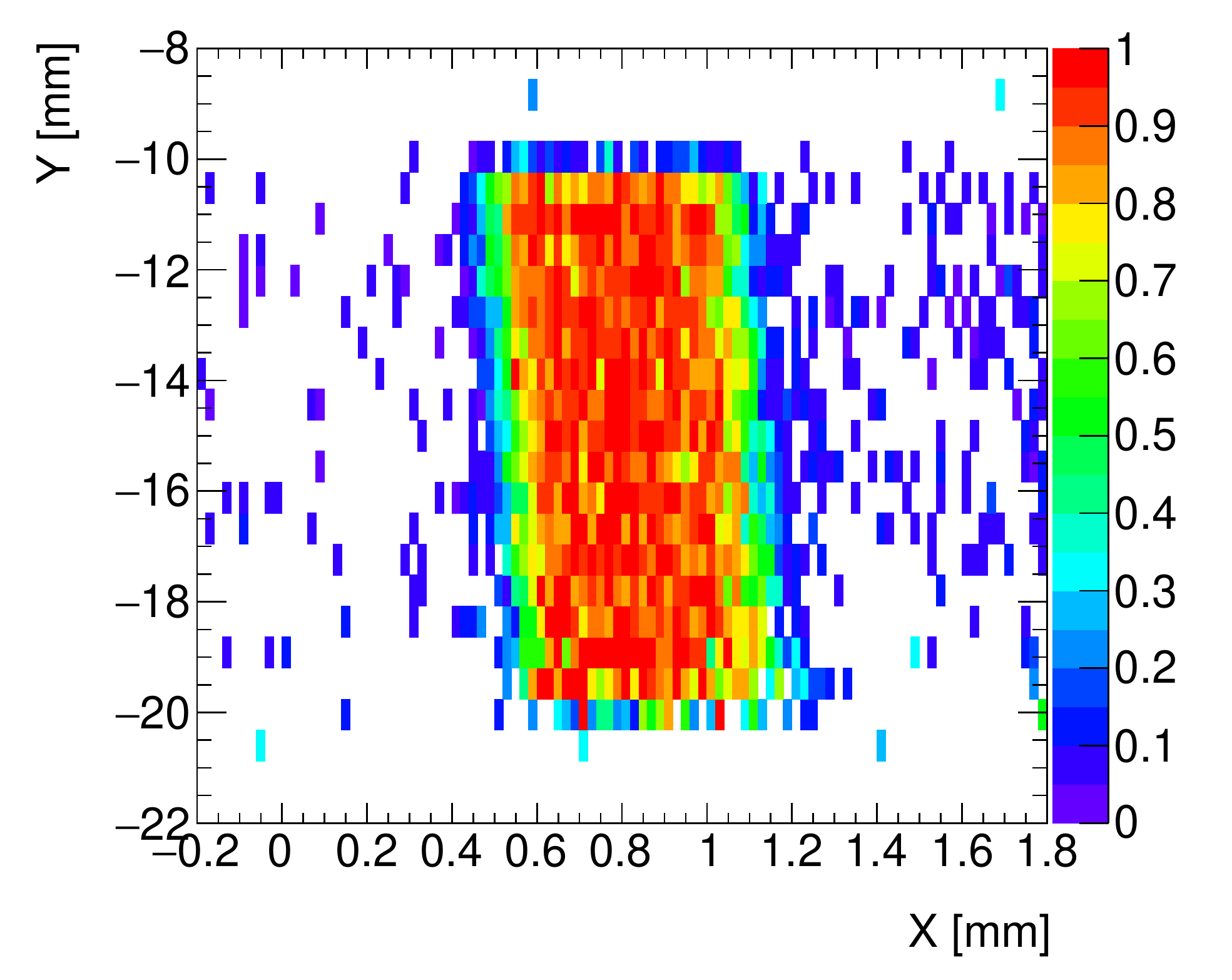}
\caption{$X$-$Y$ efficiency map of strip sensor for signal threshold of 14mV.}

\label{fig:Effstrip2D}
\end{center}
\end{figure}

\begin{figure}[h]
    \begin{center}   
 \includegraphics[width=60mm]{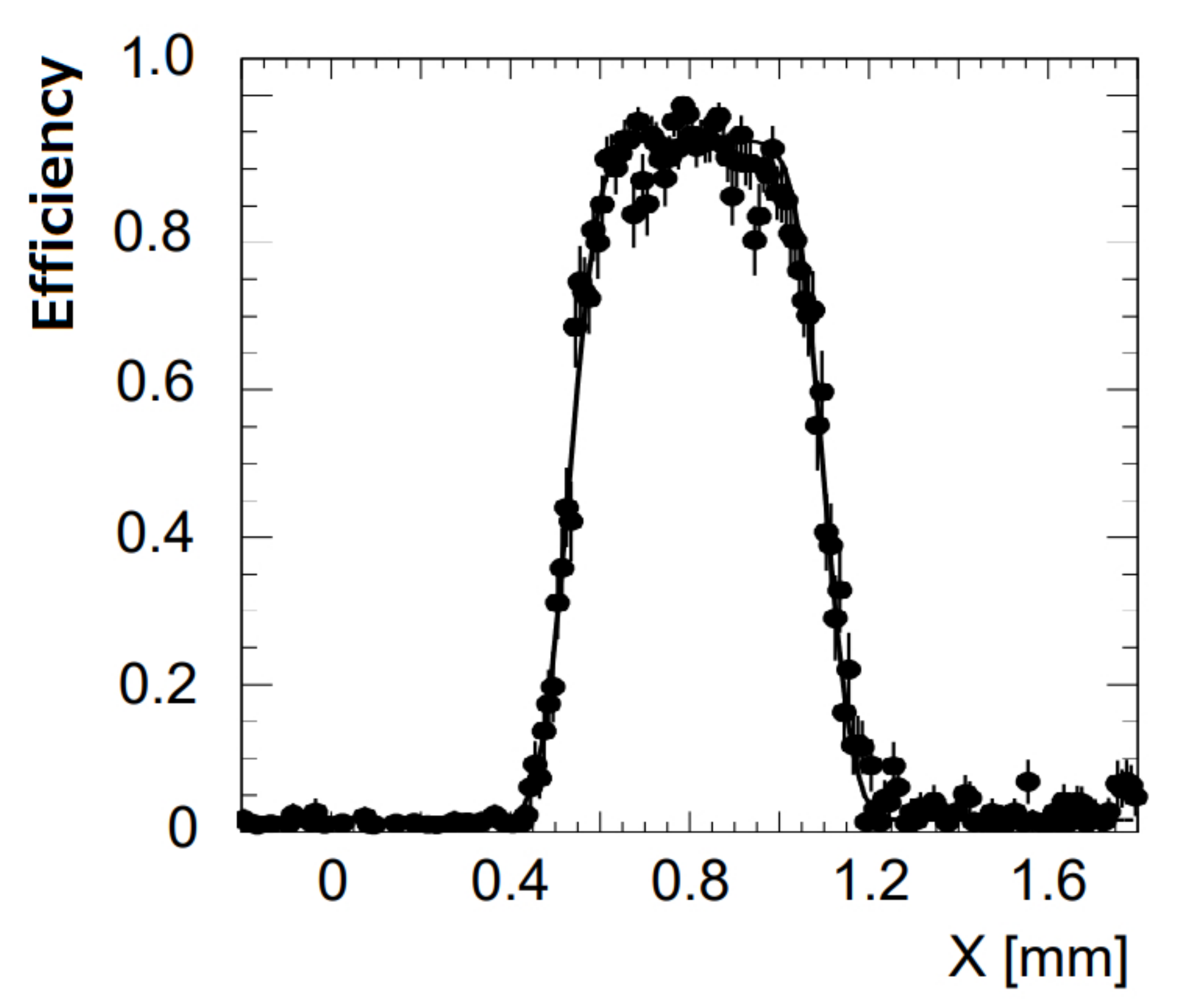}   
 \caption{$X$ projection of the efficiency map in the $y$ range from -17~mm to -12~mm, after correcting for the rotation, see text. The edge distribution is fitted to an error function.}
         \label{fig:Effstrip1D}
 \end{center}
\end{figure}

Signal efficiency is measured for the strip and pad sensors.
For the strip, consecutive 7 strips were read out at 170~V of bias voltage. 
The signal efficiency map in $x$-$y$ coordinates is shown in Fig.~\ref{fig:Effstrip2D}. The threshold to the signal amplitude was set to 14~mV.  
Fig.~\ref{fig:Effstrip1D} is the $x$ projection of Fig.~\ref{fig:Effstrip2D} for $y$ range from -17~mm to -12~mm after correcting for the rotation with respect to the strip direction: the parallelogram feature of Fig.~\ref{fig:Effstrip1D}  distribution is corrected. The efficiency in the center region is efficient and drops towards the edges due to the finite track pointing resolution affected substantially by multiple scattering.


\begin{figure}[h]
    \begin{center}
    \includegraphics[width=60mm]{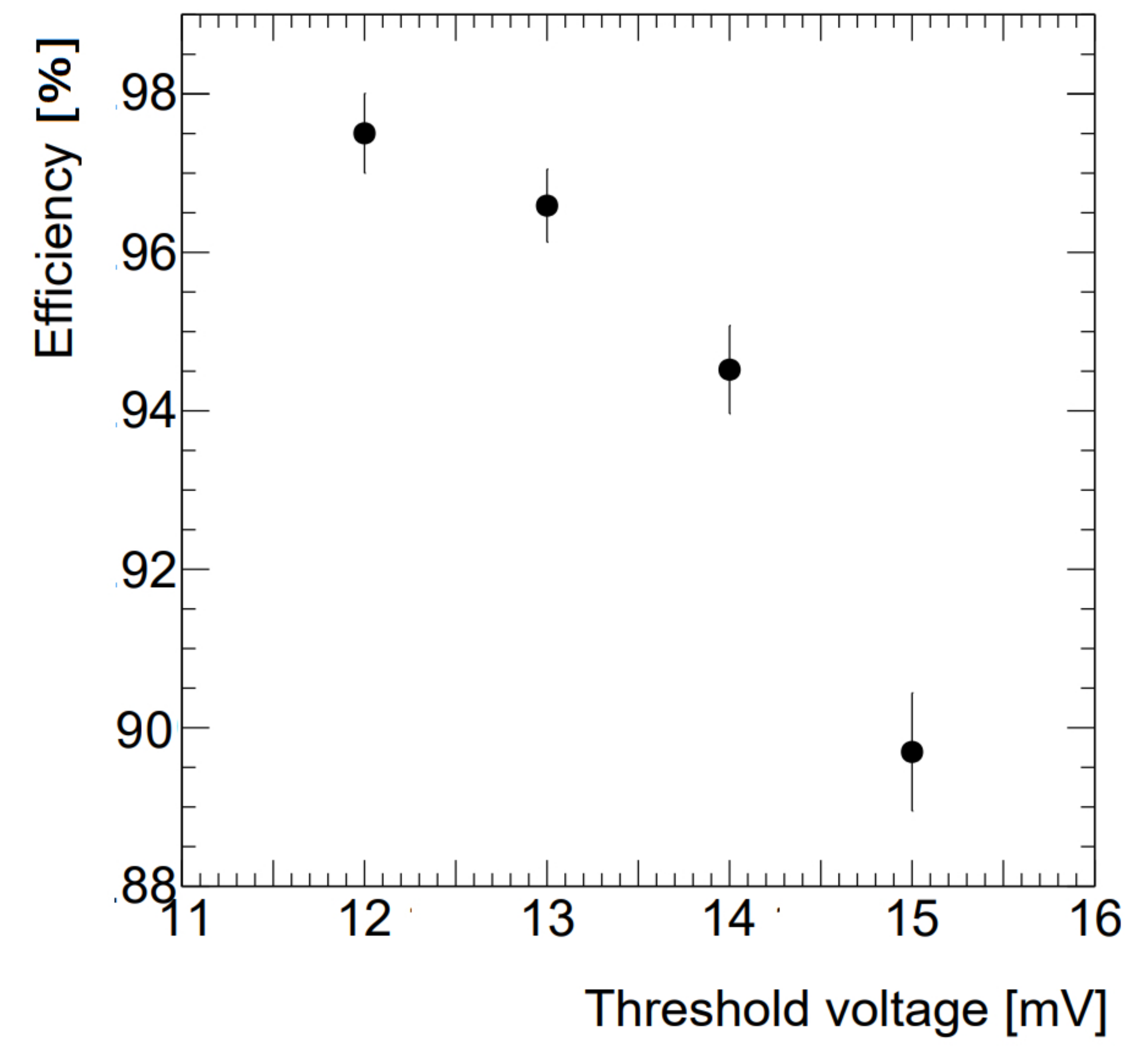}
    \caption{Signal efficiency vs the threshold to the pulse height.}
    \label{fig:EffvsThr2}
    \end{center}
\end{figure}
The signal efficiency was calculated by fitting two error functions for the distributions at edges and a constant at the center region. The efficiency defined by the constant is 94.5 $\pm$ 0.8\%. The efficiency as a function of the signal amplitude threshold is shown in Fig.~\ref{fig:EffvsThr2}.

\begin{figure}[h]
    \begin{center}
        \subfigure[upstream pad]{
           \includegraphics[width=42mm]{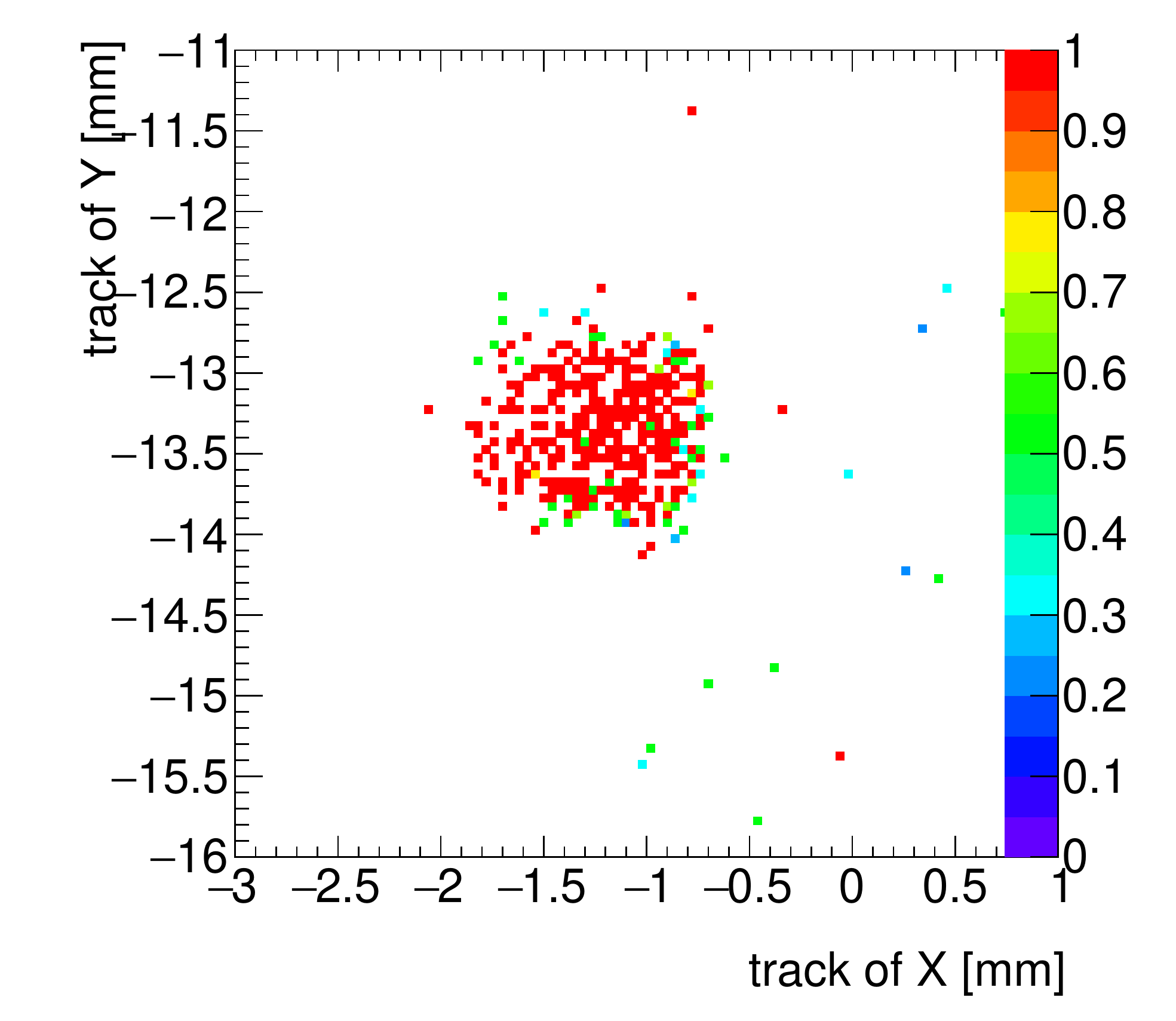}
        }
        \subfigure[downstream pad]{
            \includegraphics[width=42mm]{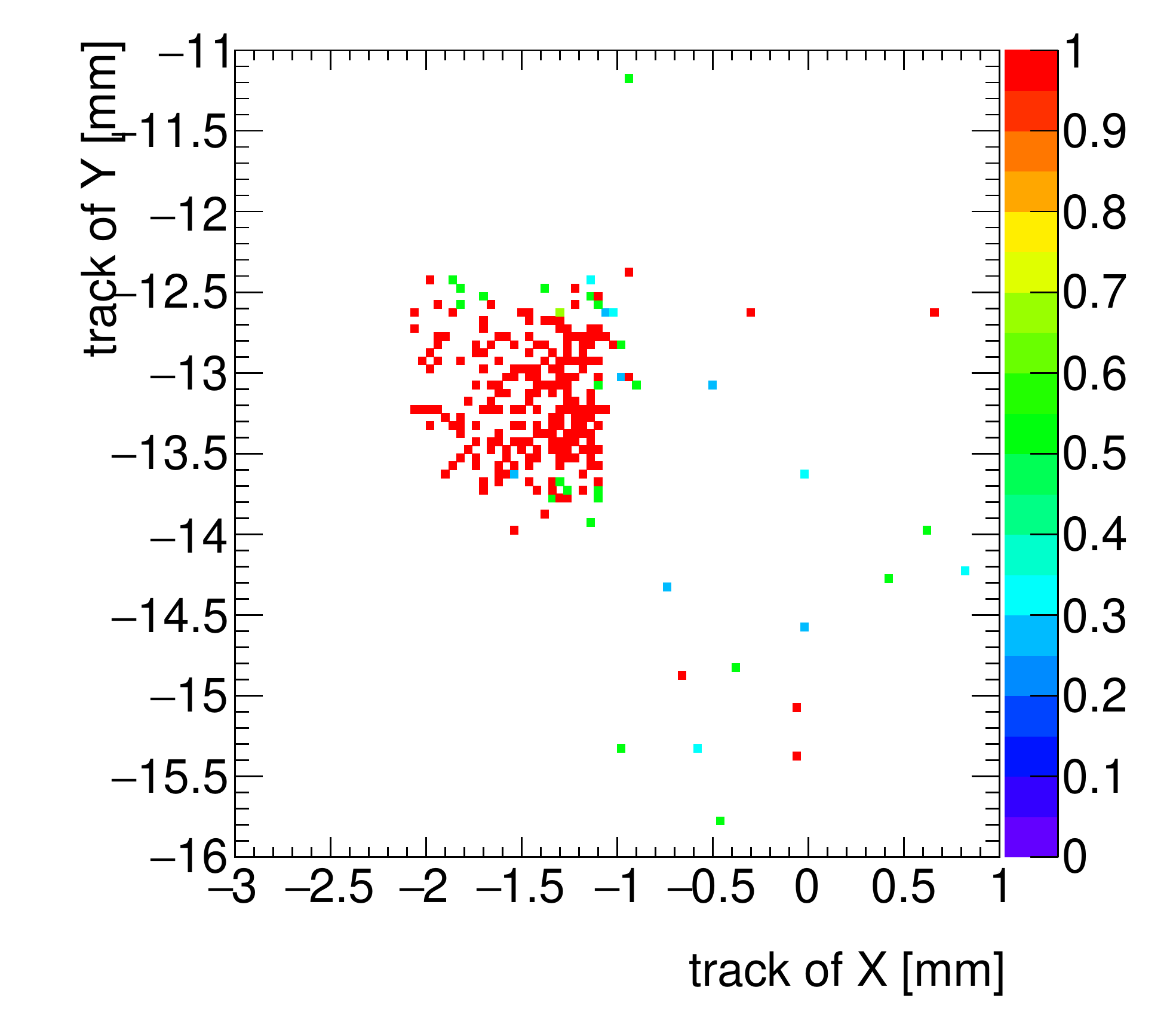}
        }
           \caption{Efficiency maps of two pad sensors stacked on the same amplifier board}
             \label{fig:Effpad}
    \end{center}
\end{figure}

The efficiency maps for two pad sensors stacked on the same amplifier board are shown in Fig.~\ref{fig:Effpad}.
To increase the statistics, the data measured at three bias voltage settings (170~V, 160~V and 140~V) are combined. The threshold to the signal amplitude is set to 18~mV. The efficiency is calculated as the average in the fiducial region which is defined as the LGAD pad active area but excluding the 3-sigma boundary with respect to the track pointing resolution.
Combining up and down stream, the pad efficiency is 97.9 $\pm$ 0.9\%.

\subsection{Spatial resolution 
}


\begin{figure}[h]
    \begin{center}
    \includegraphics[width=85mm]{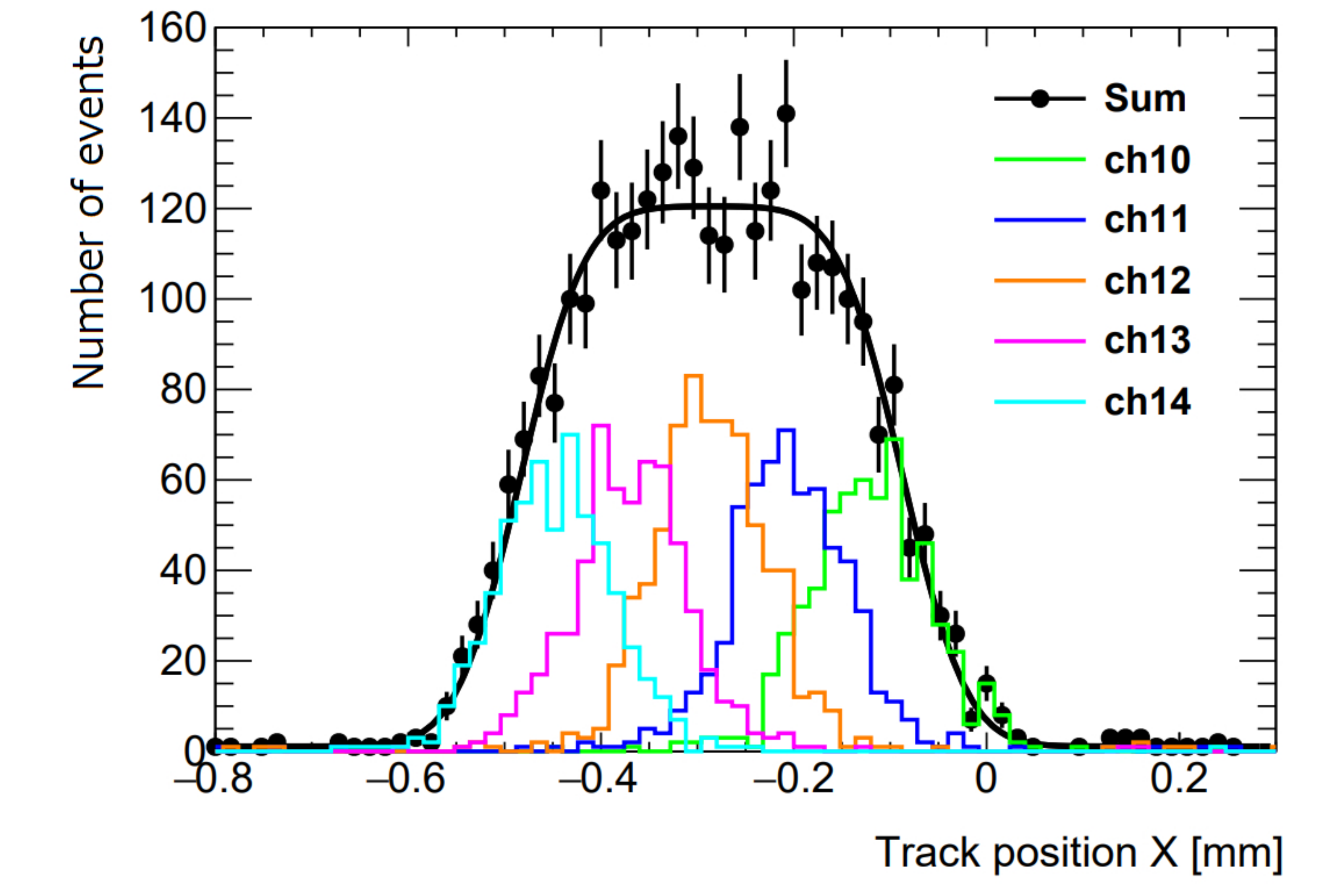}
    \caption{Hit distributions as function of the track position. The individual histograms are for the hits in consecutive strip channels, while the black curve with dots is sum of the individual histograms.}
    \label{fig:preso}
    \end{center}
\end{figure}
For 80~{\textmu}m pitch strip sensors, the spatial resolution is expected to be about 23~{\textmu}m assuming binary readout. Consecutive 7 strips were read out and 5 strips at center were used to measure the spatial resolution while the outer strips were used to suppress the cross talk signal from outer strips other than the 7 strips under read out. The same data were used for the efficiency measurement of the strip sensor. To keep more than 95\% efficiency, the threshold to the signal amplitude was set to 14~mV. 

The hit distributions of the central five strips are shown in Fig.~\ref{fig:preso} as a function of the track position.  Colored histograms are for the strips in consecutive order, and black points are sum of all histograms. The spatial resolution was calculated by fitting each histogram with a Gaussian function and subtracting the track pointing resolution. The pointing resolution was calculated from the sum distribution as it is expected to be a step function if the pointing is perfect. The obtained spatial resolution is 20.3 $\pm$  3.2~{\textmu}m; the result is consistent to the expected spatial resolution. The spatial resolution will be improved by taking charge weighted positions - as the pointing resolution is significantly large, such estimate is not 
discussed here.


\section{Radiation Tolerance 
}
\label{sec:radiationtolerance}

Radiation hardness of the developed devices were evaluated by irradiating with $^{60}$Co $\gamma$'s and protons. The $\gamma$ irradiation is to investigate the total ionization dose (TID) effect, while the proton irradiation is to investigate in addition the non-ionization energy loss (NIEL) affecting to the bulk silicon damage. The $p^+$ doping concentration is also reduced by NIEL by the process aka acceptor removal. 

\subsection{Irradiations 
}
The proton irradiation facility of Cyclotron and Radioisotope Center (CYRIC), Tohoku University, was used to irradiate the devices to evaluate the NIEL effect. High intensity proton beam with a momentum of 70~MeV at typically beam current of 300~nA provided the fluence of a few times 10$^{14}$~1-MeV n$_{\mathrm eq}$/cm$^2$ in a 20$\times$20~mm$^2$ area uniformly within an hour. The devices were irradiated in a cooling box of which temperature was maintained at -15$^\circ$C by a cold nitrogen gas flow. The irradiation system and fluence control procedure are detailed in Ref.~\cite{Nakamura2015}.

The $^{60}$Co $\gamma$-ray irradiation facility at National Institutes for Quantum Science and Technology (QST) at Takasaki, Japan served $\gamma$-rays at 0.1$\sim$10~kGy/h with a 6~PBq $^{60}$Co source ~\cite{qst}. The devices were irradiated at room temperature to the doses from 10~kGy to 1~MGy to evaluate the TID effect. The total dose was controlled by the irradiation time and the dose rate given at the sample position.
The dose rate was checked by Alanine dosimeters (Aminogray) for the first 2 hours and assumed to be stable for the entire irradiation period.  

\subsection{Acceptor removal effect 
}
As one of the NIEL effects,  electrically active shallow acceptors are transformed into defect complexes that are no longer having the properties of those shallow dopants. The effect, called acceptor removal, introduces a serious effect to the LGAD detector in general as the reduction of $p^+$ doping concentration of the gain layer increases the operation voltage due to the higher gain voltage as discussed in Section~\ref{Sec::LeakageCurrent}.

\begin{figure}[h]
    \begin{center}
    \includegraphics[width=70mm]{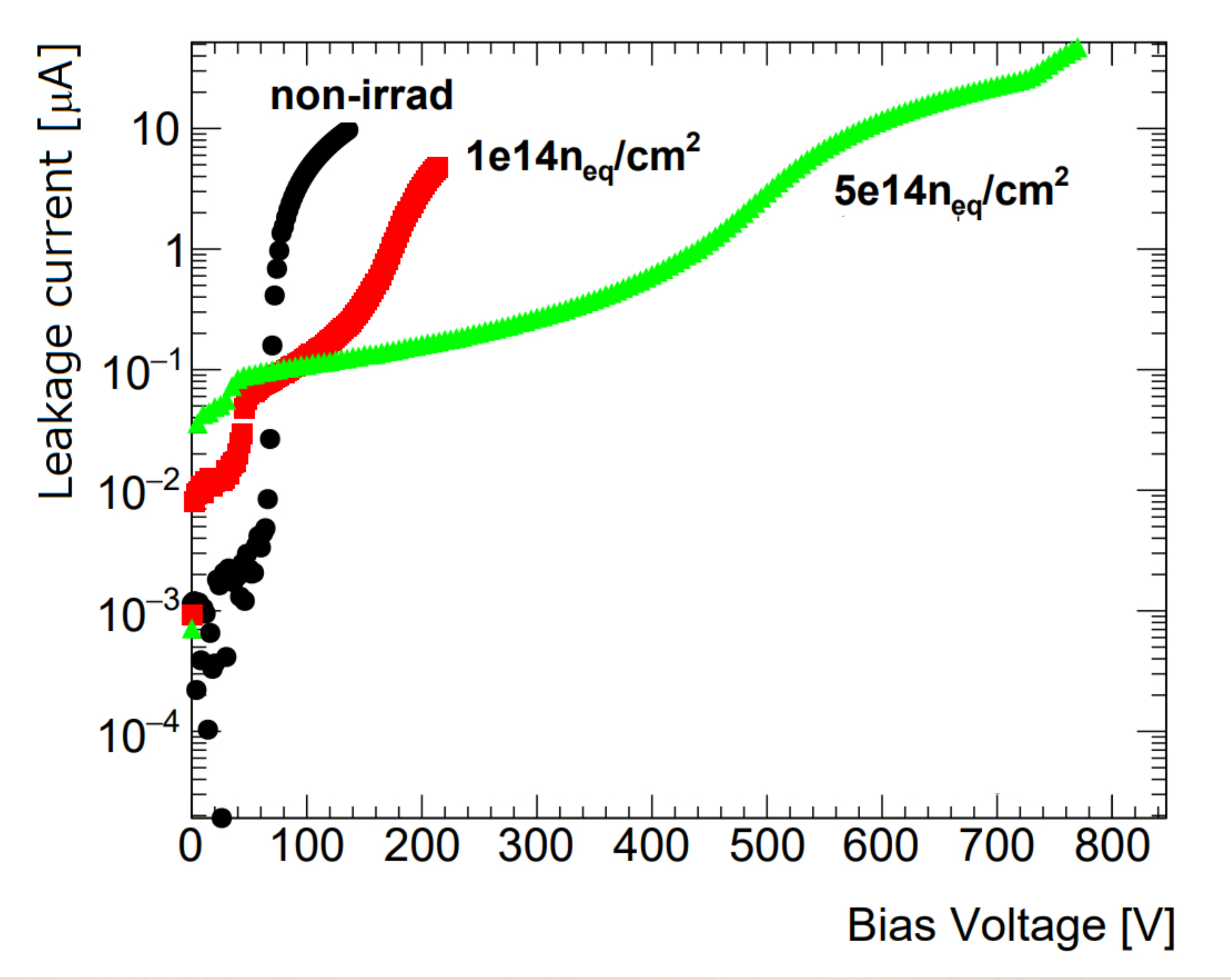}
    \caption{IV characteristics of non-irradiated and two proton irradiated B-3 pad sensors, measured at 20$^{\circ}$C for non-irradiated and at -20$^{\circ}$C for irradiated samples.}
    \label{fig:IVpadproton}
    \end{center}
\end{figure}

\begin{figure}[th]
    \begin{center}
    \includegraphics[width=80mm]{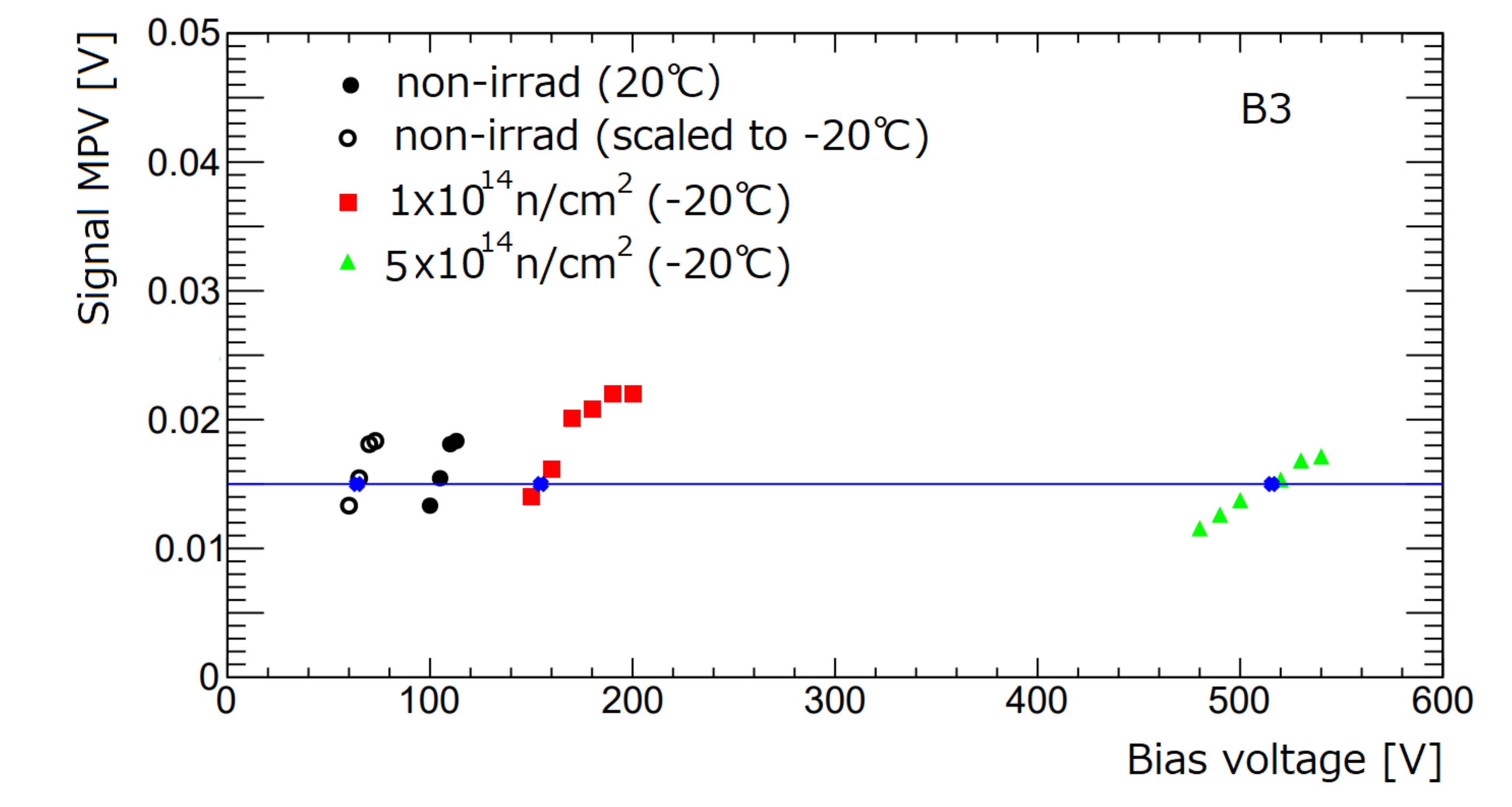}
    \caption{Signal size of proton irradiated sensors as function of the bias voltage for B3 pad type sensors.The irradiated samples were measured at -20$^{\circ}$C, while non-irradiated sample was measured at +20$^{\circ}$C with values calculated for -20$^{\circ}$C also shown.}
    \label{fig:SignalMPVbias}
    \end{center}
\end{figure}

\begin{figure}[H]
    \begin{center}
    \includegraphics[width=50mm]{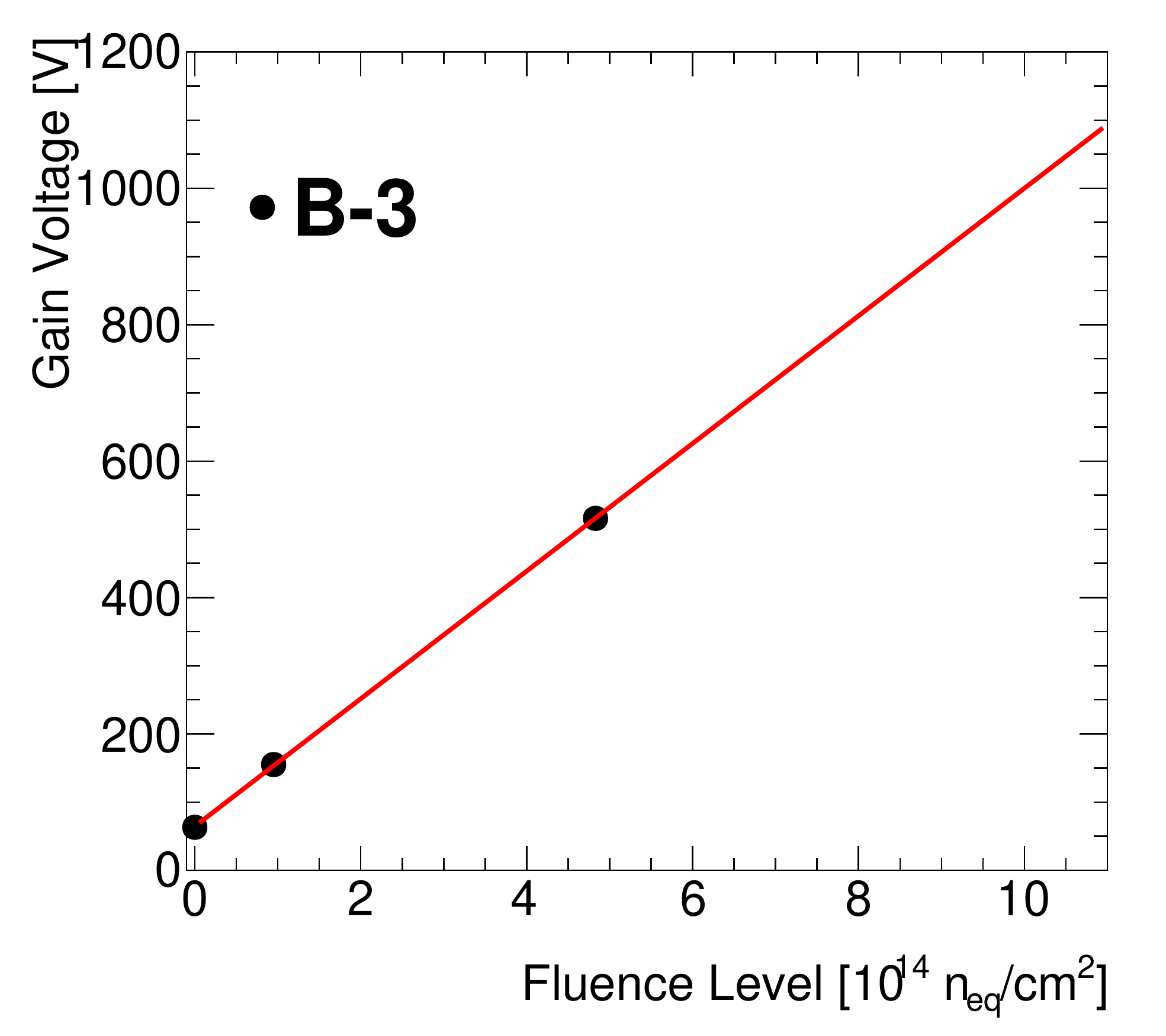}
    \caption{The gain voltage obtained of B-3 type pad by the MIP signal as a function of proton fluence. The red line shows the fit by linear function.}
    \label{fig:SignalMPVfluence}
    \end{center}
\end{figure}

The acceptor removable effect was studied for the sensors irradiated at CYRIC to 1$\times$10$^{14}$ and 5$\times$10$^{14}$~n$_{\mathrm eq}$/cm$^2$ fluences. Fig.~\ref{fig:IVpadproton} shows the leakage current as a function of the bias voltage for three B-3 pad sensors, non-irradiated,  1$\times$10$^{14}$ and 5$\times$10$^{14}$~n$_{\mathrm eq}$/cm$^2$. The tendency that the gain voltage increases with the fluence is clearly observed. To quantify the tendency, the MPVs of signal pulse height distributions are plotted as a function of bias voltage as shown in Fig.~\ref{fig:SignalMPVbias}. The irradiated samples were measured at $-20^{\circ}$C while the non-irradiated sensor was measured at $+20^\circ$C: the value was correct to that at $-20^{\circ}$C by taking the measured temperature dependence of the leakage current. This procedure was necessary since the gain voltage at -20$^{\circ}$C is lower than the full depletion voltage of the bulk. The gain voltages are obtained to be MPV of 15~mV. The gain voltage is proportional to the fluence and the slope is evaluated as 96~V/10$^{14}$~n$_{\mathrm{eq}}$/cm$^2$ by fitting a linear function, as shown in Fig.~\ref{fig:SignalMPVfluence}.
This indicates that the gain voltage at 1$\times$10$^{15}$~n$_{\mathrm{eq}}$/cm$^2$ is approximately 1~kV, which exceeds the bias tolerance of the current sensor design.

\subsection{Total ionization dose effect 
} 
\begin{figure}[h]
    \begin{center}
    \includegraphics[width=70mm]{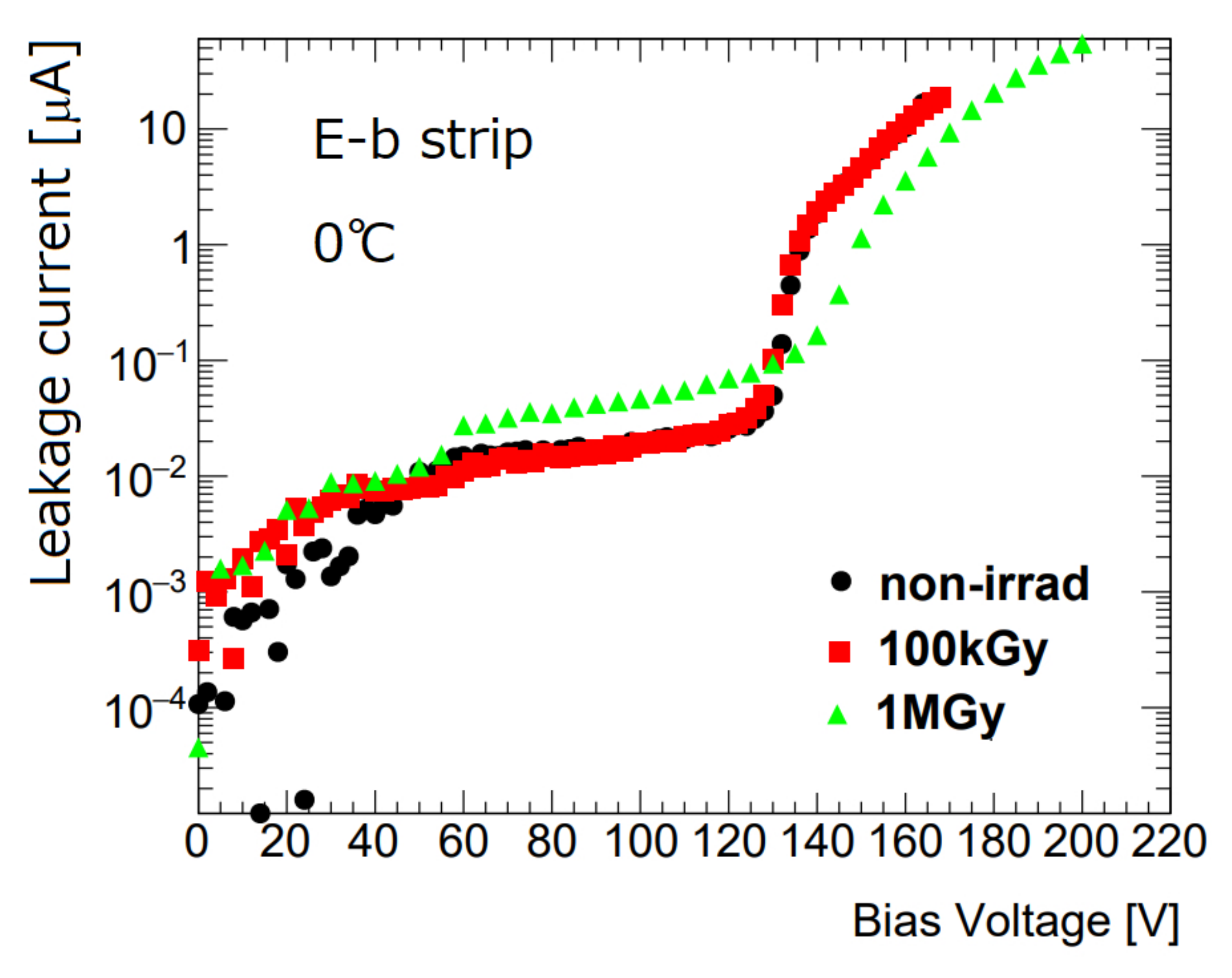}
    \caption{IV characteristic for the strip samples irradiated with $\gamma$s (non-irrad, 100~kGy, 1~MGy)}
    \label{fig:IVgamma}
    \end{center}
\end{figure}

The major consequence of the TID damage is the surface damage, trapping positive charges at the Si and SiO$_2$ interface and in  SiO$_2$ layer.
The magnitude of cross talk is sensitive to the trapped charges. The TID effect was evaluated for the E-b type strip samples irradiated with $\gamma$-rays to dose of 100~kGy and 1~MGy.  Fig.~\ref{fig:IVgamma} shows the $IV$ characteristics. Significant irradiation-induced leakage current increase as in the NIEL damage was not observed, as NIEL  is significantly small in $\gamma$ irradiation.  

\begin{figure}[htbp]
    \begin{center}
    \includegraphics[width=75mm]{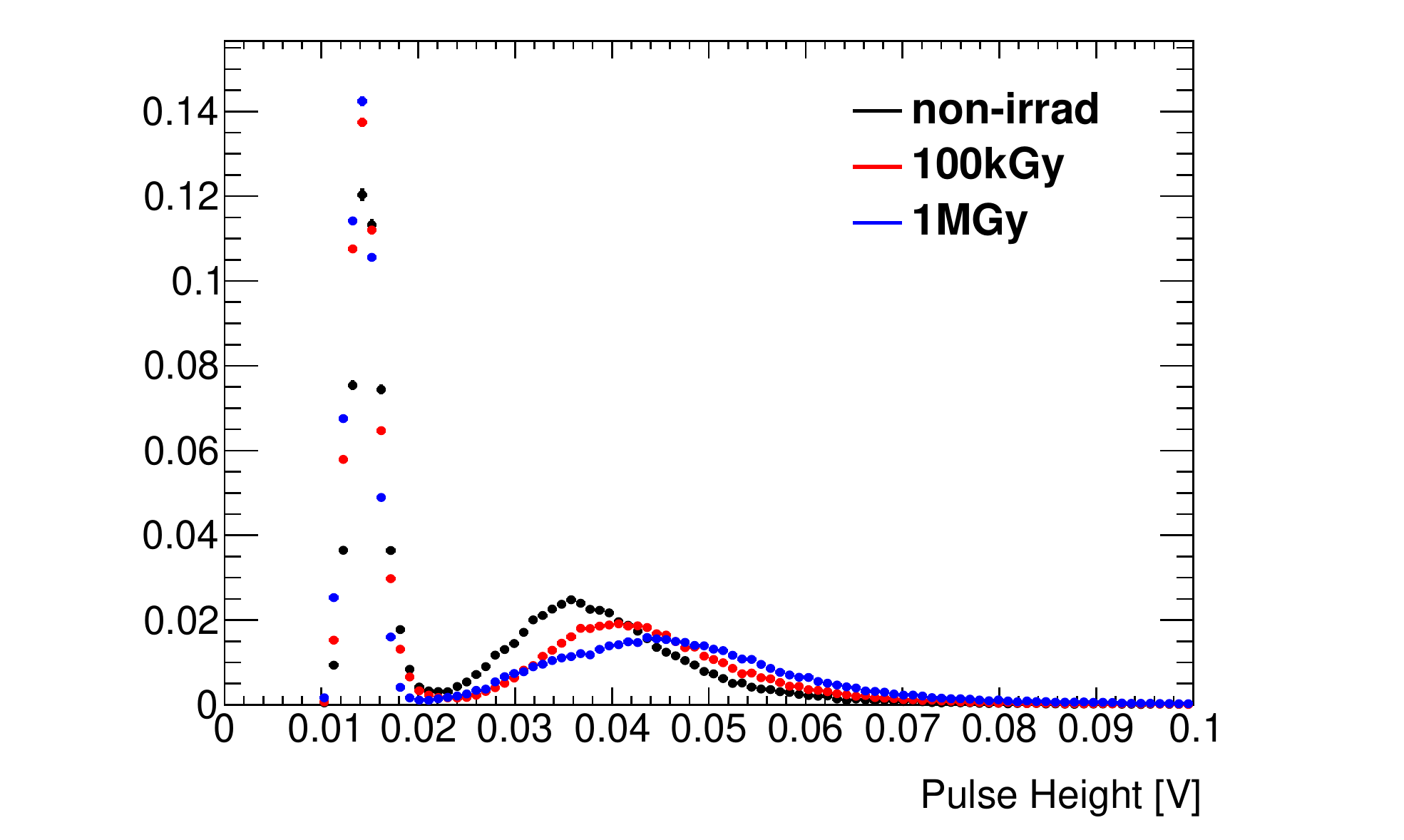}
    \caption{Pulse Height distributions summed over the leading and second leading strips for non-irradiated, 100~kGy and 1~MGy irradiated samples.}
    \label{fig:sumPulse}
    \end{center}
\end{figure}

The distributions of charge sum of the leading and second-leading strips measured with $\beta$'s are plotted in Fig.~\ref{fig:sumPulse}. The charge collection was not degraded, or rather increased by a factor of 1.2 for the sample irradiated to 1~MGy. The reason is presented in the end of this section.



\begin{figure}[htbp]
    \begin{center}
        \subfigure[Non-irradiated sample]{
        \includegraphics[width=42mm]{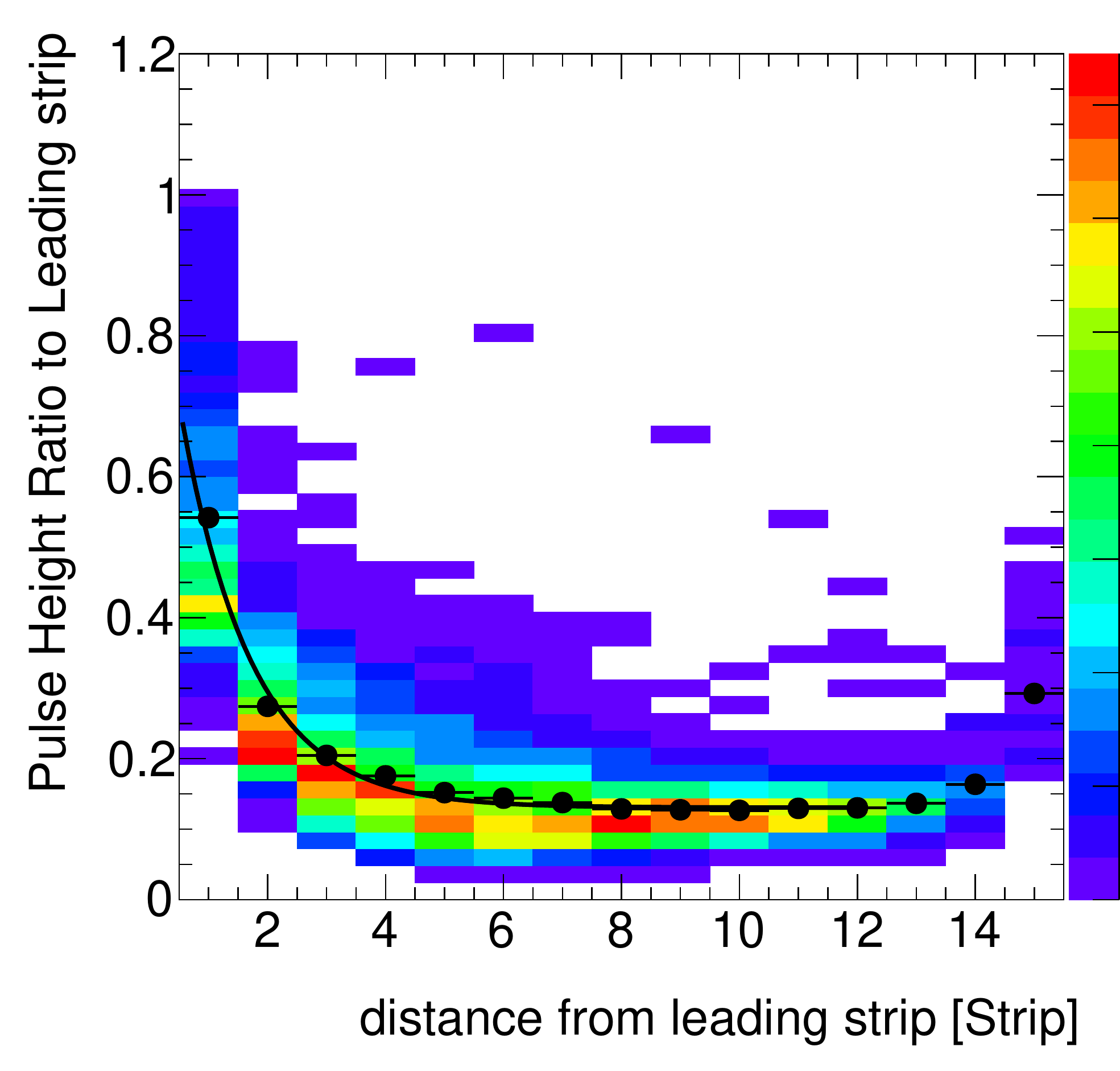}
        \label{fig:crosstalk_nonirrad}
        }
       \subfigure[100~kGy irradiated sample.]{
        \includegraphics[width=42mm]{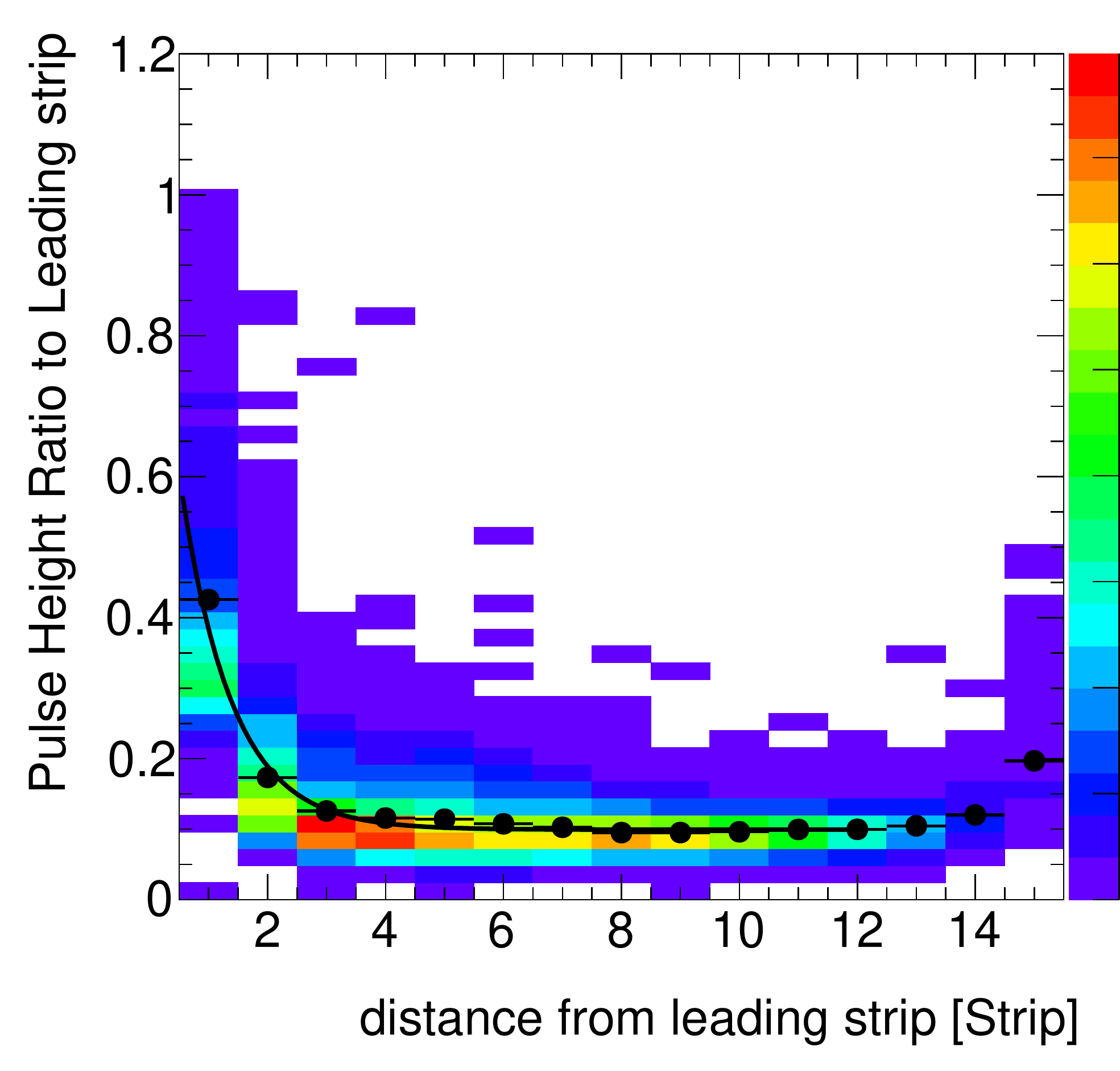}
        \label{fig:crosstalk_irrad100k}
        }
         \subfigure[1~MGy irradiated sample.]{
            \includegraphics[width=42mm]{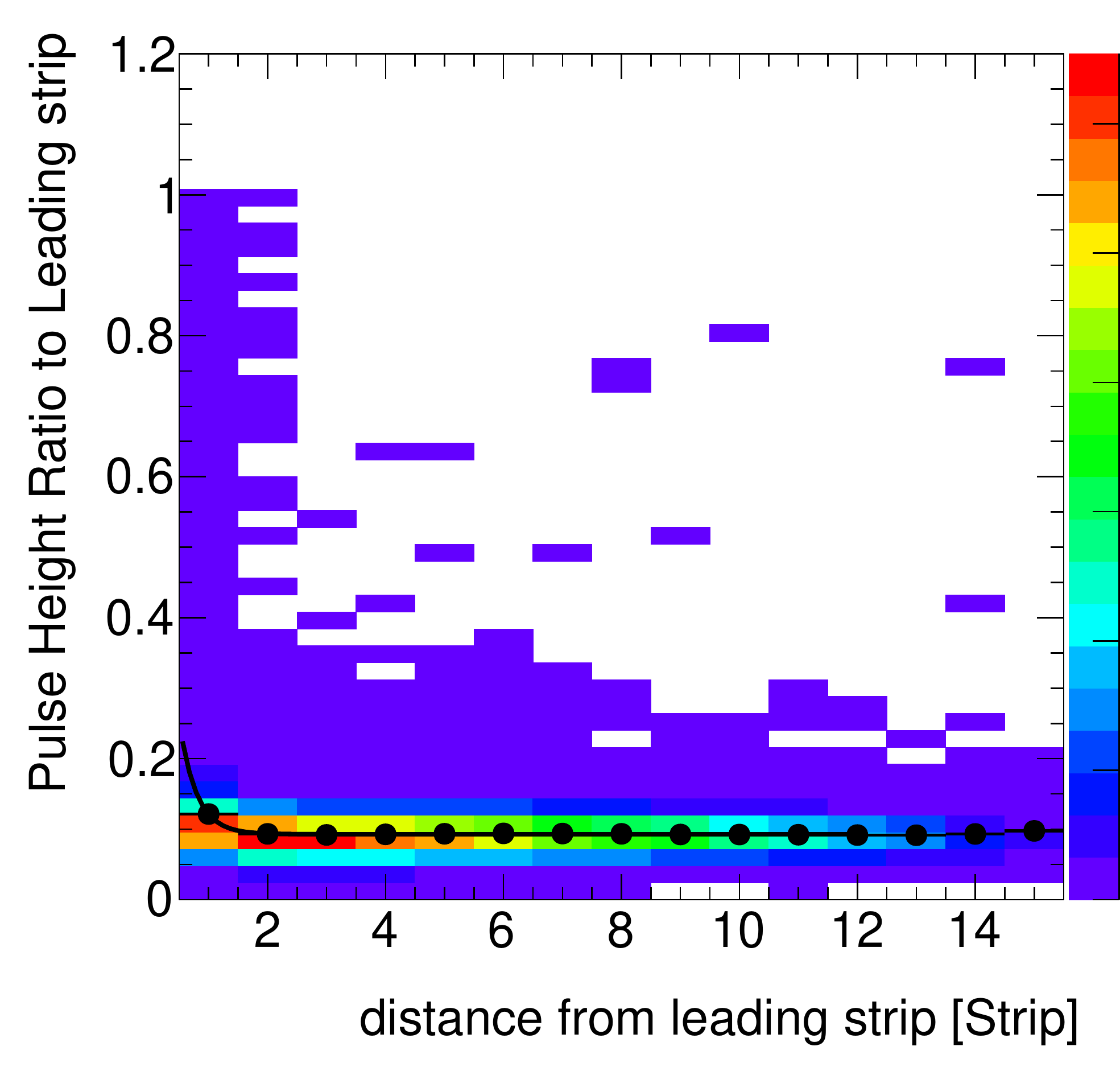}
            \label{fig:crosstalk_irrad1M}
        }
        \caption{Pulse height ratio to the leading strip as function of distance from the leading strip. Unit is in number of strips (80~$\mu$m pitch) }\label{fig:crosstalk_irrad}
    \end{center}
\end{figure}

The magnitude of the cross talk in the 80~{\textmu}m-pitch strip sensors is evaluated. The pulse height ratio to the leading channel pulse height as a function of distance from the leading strip  is shown in Fig.~\ref{fig:crosstalk_irrad} for non-irradited, 100~kGy and 1~MGy irradiated samples. 
The cross talk distance for non-irradiated, 100~kGy and 1~MGy are
1.20$\pm$0.02, 
0.87$\pm$0.01, and
0.29$\pm$0.01 strips,
respectively. 
 The result indicates that the positive trapped charges at the surface reduces the cross talk, as explained that flow of electrons through the $n^+$ implant layer is disturbed by the trapped positive charges. The phenomena clarify the observation of the charge sum in the leading two strips increases after TID damage; the charge sum of all strips showed to remain unchanged. 

\section{Application for visible light detection 
}
The fast timing detector with spatial resolution has possibility of application in various fields other than high-energy physics experiments.  Such devices capable of detect infra-red and/or visible light, for example, will be unique for bio and medical sciences.

The detectors described in this paper have aluminum electrodes on the top surface. To avoid the light reflection, an AC-LGAD detector with transparent electrodes has been developed.

High conductivity of the AC electrode on the oxide layer is important to keep the impedance small such that the signal is efficiently picked up and transported out to the amplifier. Therefore conductive and transparent electrodes are necessary. As a trial, thin poly-Si was tested. While the conductivity of poly-Si can be varied by doping, poly-Si with a sheet resistivity of  6.5~k$\Omega$/$\square$ was examined as a first prototype.

The signal which AC-LGAD with transparent poly-Si electrodes detected visible red light, infra-red light and beta rays ($^{90}$ Sr) as shown in the Fig. 25.
The beta ray signal was recorded by the Scintilator trigger with the same setup described in Sec.~\ref{Measurementsetup}. Infra-red and red light laser system were used to evaluate the pulse height distribution. As a result, the signal induced by infra-red and red laser have been 
observed. Further studies like transmittance of the poly-Si electrode and attenuation of signal in the electrodes will be performed.

\begin{figure}[htbp]
    \begin{center}
    \includegraphics[width=75mm]{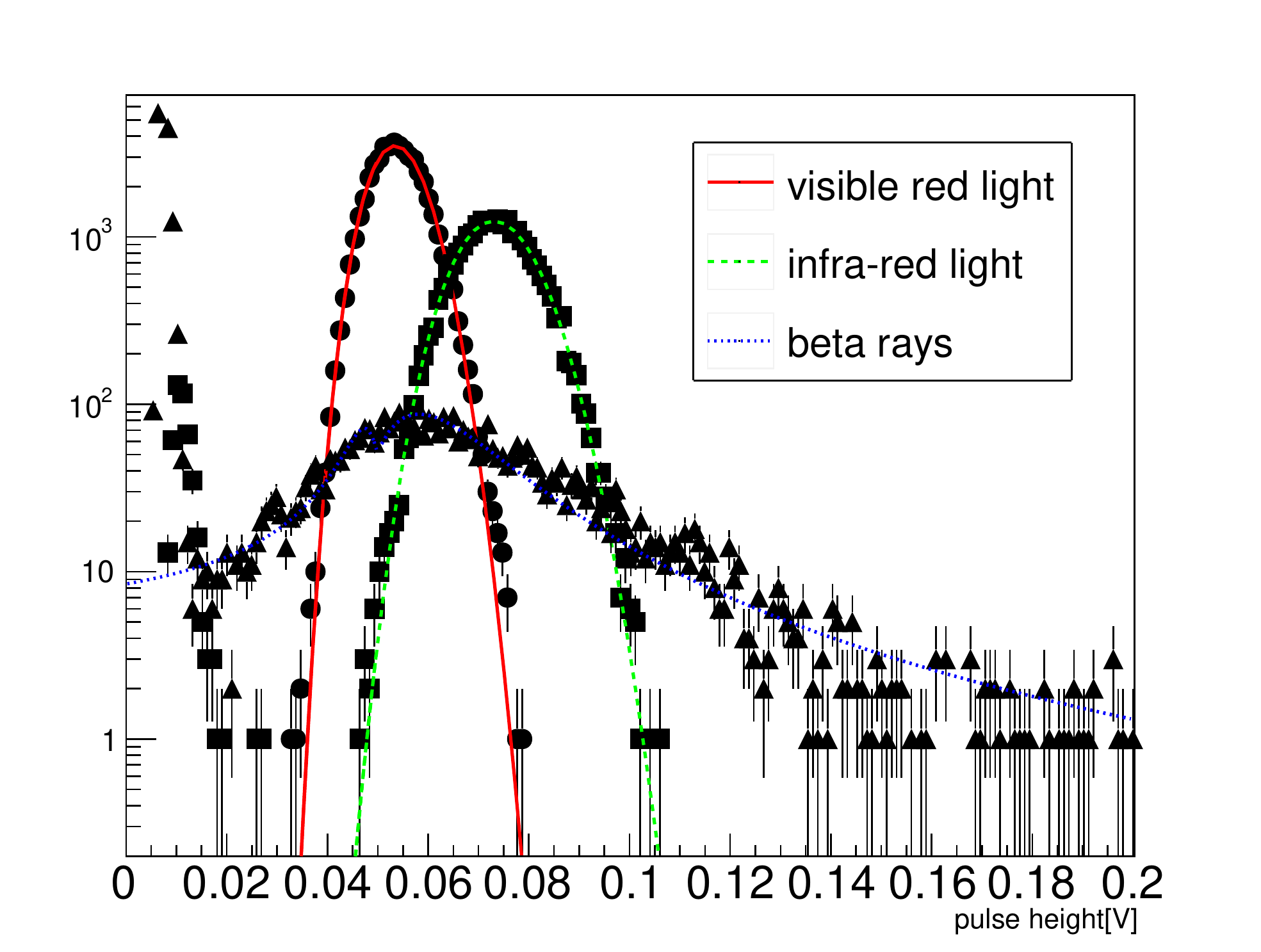}
    \caption{Pulse Height distributions of poly-Si electrode AC-LGAD detector for beta ray (blue), infra-red light (green) and visible red light (red).}
    \label{fig:poly}
    \end{center}
\end{figure}


\section{Conclusions 
}
Development  of capacitive-coupled Low-Gain Avalanche Diode (AC-LGAD) sensors for precise time and spatial resolution has been in progress with Hamamatsu Photonics. The three process parameters, $n^+$ and $p^+$ implant concentrations and coupling capacitance, were varied and the fundamental AC-LGAD performance was evaluated to optimize the parameters.

From the signal size and cross talk size characterization of 80~{\textmu}m pitch strip type sensors, lower $n^+$ doping concentration results in larger signal size and smaller cross talk, and the sample with 1600$\Omega$/$\square$ $n^+$ resistivity (E-type) exhibits the best performance.
Further increase of the resistivity than this value is not possible due to early breakdown. 

Strip sensors were tested and 99.98\% efficiency at the 10$^{-4}$ noise rate was obtained.
The spatial resolution of 20.3$\pm$3.2~{\textmu}m was obtained for the 80~{\textmu}m pitch strip sensor with binary readout. A charge weighting position evaluation is also possible with relatively low $n^+$ resistivity sensor and C-2 pad sensor achieved a spatial resolution of 6-10~{\textmu}m~ and 30~ps time resolution simultaneously \cite{Heller_2022}. 

The time resolution above was evaluated using 120~GeV proton beam.  A test system to measure time resolution using a $^{90}$Sr $\beta$-ray source is equipped. A resolution of 44.2$\pm$1.5~ps and 43.9$\pm$1.9~ps were obtained for E-b and C-2 type respectively, which is worse than the beam result but useful for the detector development.  

AC-LGADs with transparent poly-Si electrodes were fabricated for visible and infra-red light detection. The beta-ray signals (MIPs) were observed similar to the Aluminum electrode AC-LGAD as well as infra-red and red light signals were detected. The results indicate that the AC-LGAD device has a possibility to use as light detection sensors for the application in various fields.

 
\section*{Acknowledgments}
This research was partially supported by Grant-in-Aid for scientific research on advanced basic research (Grant No. 19H05193, 19H04393, 21H0073 and 21H01099) from the Ministry of Education, Culture, Sports, Science and Technology, of Japan as well as the Proposals for the U.S.-Japan Science and Technology Cooperation Program in High Energy Physics from JFY2019 to JFY2023 granted by High Energy Accelerator Research Organization (KEK) and Fermi National Accelerator Laboratory (FNAL). In keeping the research program move forward, the following facilities were very important to this project :  Research Center for ELectron PHoton Science (ELPH) at Tohoku University, National Institutes for Quantum Science and Technology (QST) at Takasaki, Cyclotron Radio Isotope Center (CYRIC) at Tohoku University, and Fermilab Test Beam Facility (FTBF) at Fermi National Accelerator Laboratory (FNAL).




\bibliographystyle{elsarticle-num-names} 
\bibliography{ACLGAD2021.bib}

\begin{thebibliography}{12}
\expandafter\ifx\csname natexlab\endcsname\relax\def\natexlab#1{#1}\fi
\providecommand{\url}[1]{\texttt{#1}}
\providecommand{\href}[2]{#2}
\providecommand{\path}[1]{#1}
\providecommand{\DOIprefix}{doi:}
\providecommand{\ArXivprefix}{arXiv:}
\providecommand{\URLprefix}{URL: }
\providecommand{\Pubmedprefix}{pmid:}
\providecommand{\doi}[1]{\href{http://dx.doi.org/#1}{\path{#1}}}
\providecommand{\Pubmed}[1]{\href{pmid:#1}{\path{#1}}}
\providecommand{\bibinfo}[2]{#2}
\ifx\xfnm\relax \def\xfnm[#1]{\unskip,\space#1}\fi
\bibitem[{CER(2020)}]{CERN-LHCC-2020-007}
\bibinfo{title}{{Technical Design Report: A High-Granularity Timing Detector
  for the ATLAS Phase-II Upgrade}}, \bibinfo{type}{Technical Report}, CERN,
  \bibinfo{address}{Geneva}, \bibinfo{year}{2020}. \URLprefix
  \url{https://cds.cern.ch/record/2719855}.
\bibitem[{Yang et~al.(2020)Yang, Alderweireldt, Atanov, Ayoub, {da Costa},
  García, Chen, Christie, Cindro, Cui, D’Amen, Davydov, Fan, Galloway, Ge,
  Gee, Giacomini, Gkougkousis, Grieco, Grinstein, Grosse-Knetter, Guindon, Han,
  Howard, Huang, Jin, Jing, Kiuchi, Kramberger, Kuwertz, Labitan, Lange, Leite,
  Li, Li, Liu, Liu, Liu, Liang, Liang, Lockerby, Lyu, Mandić,
  Martinez-Mckinney, Mazza, Mikuž, Padilla, Qi, Quadt, Ran, Ren, Rizzi, Rossi,
  Sadrozinski, Saito, Schumm, Schwickardi, Seiden, Shan, Shi, Shi, Ferreira,
  Sun, Tan, Tricoli, Wan, Wilder, Wu, Wyatt, Xiao, Yang, Yang, Yu, Zhao, Zhao,
  Zhao, Zhao, Zheng, and Zhuang}]{YANG2020164379}
\bibinfo{author}{X.~Yang}, \bibinfo{author}{S.~Alderweireldt},
  \bibinfo{author}{N.~Atanov}, \bibinfo{author}{M.~Ayoub},
  \bibinfo{author}{J.~B.~G. {da Costa}}, \bibinfo{author}{L.~C. García},
  \bibinfo{author}{H.~Chen}, \bibinfo{author}{S.~Christie},
  \bibinfo{author}{V.~Cindro}, \bibinfo{author}{H.~Cui},
  \bibinfo{author}{G.~D’Amen}, \bibinfo{author}{Y.~Davydov},
  \bibinfo{author}{Y.~Fan}, \bibinfo{author}{Z.~Galloway},
  \bibinfo{author}{J.~Ge}, \bibinfo{author}{C.~Gee},
  \bibinfo{author}{G.~Giacomini}, \bibinfo{author}{E.~Gkougkousis},
  \bibinfo{author}{C.~Grieco}, \bibinfo{author}{S.~Grinstein},
  \bibinfo{author}{J.~Grosse-Knetter}, \bibinfo{author}{S.~Guindon},
  \bibinfo{author}{S.~Han}, \bibinfo{author}{A.~Howard},
  \bibinfo{author}{Y.~Huang}, \bibinfo{author}{Y.~Jin},
  \bibinfo{author}{M.~Jing}, \bibinfo{author}{R.~Kiuchi},
  \bibinfo{author}{G.~Kramberger}, \bibinfo{author}{E.~Kuwertz},
  \bibinfo{author}{C.~Labitan}, \bibinfo{author}{J.~Lange},
  \bibinfo{author}{M.~Leite}, \bibinfo{author}{C.~Li}, \bibinfo{author}{Q.~Li},
  \bibinfo{author}{B.~Liu}, \bibinfo{author}{J.~Liu}, \bibinfo{author}{Y.~Liu},
  \bibinfo{author}{H.~Liang}, \bibinfo{author}{Z.~Liang},
  \bibinfo{author}{M.~Lockerby}, \bibinfo{author}{F.~Lyu},
  \bibinfo{author}{I.~Mandić}, \bibinfo{author}{F.~Martinez-Mckinney},
  \bibinfo{author}{S.~Mazza}, \bibinfo{author}{M.~Mikuž},
  \bibinfo{author}{R.~Padilla}, \bibinfo{author}{B.~Qi},
  \bibinfo{author}{A.~Quadt}, \bibinfo{author}{K.~Ran},
  \bibinfo{author}{H.~Ren}, \bibinfo{author}{C.~Rizzi},
  \bibinfo{author}{E.~Rossi}, \bibinfo{author}{H.-W. Sadrozinski},
  \bibinfo{author}{G.~Saito}, \bibinfo{author}{B.~Schumm},
  \bibinfo{author}{M.~Schwickardi}, \bibinfo{author}{A.~Seiden},
  \bibinfo{author}{L.~Shan}, \bibinfo{author}{L.~Shi},
  \bibinfo{author}{X.~Shi}, \bibinfo{author}{A.~S.~C. Ferreira},
  \bibinfo{author}{Y.~Sun}, \bibinfo{author}{Y.~Tan},
  \bibinfo{author}{A.~Tricoli}, \bibinfo{author}{G.~Wan},
  \bibinfo{author}{M.~Wilder}, \bibinfo{author}{K.~Wu},
  \bibinfo{author}{W.~Wyatt}, \bibinfo{author}{S.~Xiao},
  \bibinfo{author}{T.~Yang}, \bibinfo{author}{Y.~Yang},
  \bibinfo{author}{C.~Yu}, \bibinfo{author}{L.~Zhao},
  \bibinfo{author}{M.~Zhao}, \bibinfo{author}{Y.~Zhao},
  \bibinfo{author}{Z.~Zhao}, \bibinfo{author}{X.~Zheng},
  \bibinfo{author}{X.~Zhuang},
\newblock \bibinfo{title}{{Layout and performance of HPK prototype LGAD sensors
  for the High-Granularity Timing Detector}},
\newblock \bibinfo{journal}{Nucl. Instr. and Methods} \bibinfo{volume}{A980}
  (\bibinfo{year}{2020}) \bibinfo{pages}{164379}.
  \DOIprefix\doi{https://doi.org/10.1016/j.nima.2020.164379}.
\bibitem[{Onaru et~al.(2021)Onaru, Hara, Harada, Wada, KN, and
  Unno}]{HPKDCLGADRef1}
\bibinfo{author}{K.~Onaru}, \bibinfo{author}{K.~Hara},
  \bibinfo{author}{D.~Harada}, \bibinfo{author}{S.~Wada}, \bibinfo{author}{KN},
  \bibinfo{author}{Y.~Unno},
\newblock \bibinfo{title}{{Study of time resolution of low-gain avalanche
  detectors}},
\newblock \bibinfo{journal}{Nucl. Instr. and Methods} \bibinfo{volume}{A985}
  (\bibinfo{year}{2021}) \bibinfo{pages}{164664}.
  \DOIprefix\doi{https://doi.org/10.1016/j.nima.2020.164664}.
\bibitem[{Wada et~al.(2020)Wada, Onaru, Hara, Unno, and
  Nakamura}]{HPKDCLGADRef2}
\bibinfo{author}{S.~Wada}, \bibinfo{author}{K.~Onaru},
  \bibinfo{author}{K.~Hara}, \bibinfo{author}{Y.~Unno},
  \bibinfo{author}{K.~Nakamura},
\newblock \bibinfo{title}{{Design of a Segmented LGAD Sensor for the
  Development of a 4-D Tracking Detector}},
\newblock \bibinfo{journal}{PoS} \bibinfo{volume}{Vertex2019}
  (\bibinfo{year}{2020}) \bibinfo{pages}{057}.
  \DOIprefix\doi{10.22323/1.373.0057}.
\bibitem[{Nakamura et~al.(2020)Nakamura, Kita, Ueda, Hara, and
  Suzuki}]{HPKACLGADRef}
\bibinfo{author}{K.~Nakamura}, \bibinfo{author}{S.~Kita},
  \bibinfo{author}{T.~Ueda}, \bibinfo{author}{K.~Hara},
  \bibinfo{author}{H.~Suzuki},
\newblock \bibinfo{title}{{First Prototype of Finely Segmented HPK AC-LGAD
  Detectors}},
\newblock \bibinfo{journal}{JPS Conf. Proc.}  (\bibinfo{year}{2020})
  \bibinfo{pages}{010016}. \DOIprefix\doi{10.7566/JPSCP.34.010016}.
\bibitem[{eic(2022)}]{eic}
\bibinfo{title}{{The Electron-Ion Collider}}, \bibinfo{year}{2022}.
  \bibinfo{note}{\url{https://www.bnl.gov/eic/}}.
\bibitem[{Heller et~al.(2022)Heller, Madrid, Apresyan, Brooks, Chen,
  D{\textquotesingle}Amen, Giacomini, Goya, Hara, Kita, Los, Molnar, Nakamura,
  Pe{\~{n}}a, Mart{\'{\i}}n, Tricoli, Ueda, and Xie}]{Heller_2022}
\bibinfo{author}{R.~Heller}, \bibinfo{author}{C.~Madrid},
  \bibinfo{author}{A.~Apresyan}, \bibinfo{author}{W.~Brooks},
  \bibinfo{author}{W.~Chen}, \bibinfo{author}{G.~D{\textquotesingle}Amen},
  \bibinfo{author}{G.~Giacomini}, \bibinfo{author}{I.~Goya},
  \bibinfo{author}{K.~Hara}, \bibinfo{author}{S.~Kita},
  \bibinfo{author}{S.~Los}, \bibinfo{author}{A.~Molnar},
  \bibinfo{author}{K.~Nakamura}, \bibinfo{author}{C.~Pe{\~{n}}a},
  \bibinfo{author}{C.~S. Mart{\'{\i}}n}, \bibinfo{author}{A.~Tricoli},
  \bibinfo{author}{T.~Ueda}, \bibinfo{author}{S.~Xie},
\newblock \bibinfo{title}{Characterization of {BNL} and {HPK} {AC}-{LGAD}
  sensors with a 120 {GeV} proton beam},
\newblock \bibinfo{journal}{JINST} \bibinfo{volume}{17} (\bibinfo{year}{2022})
  \bibinfo{pages}{P05001}. \DOIprefix\doi{10.1088/1748-0221/17/05/p05001}.
\bibitem[{HPK(url)}]{HPK_MPPC}
\bibinfo{title}{{Hamamatsu Photonics, S13360 series}}, \bibinfo{year}{url=}.
  \bibinfo{note}{\url{https://www.hamamatsu.com/resources/pdf/ssd/s13360\_series\_kapd1052e.pdf}}.
\bibitem[{ELP(url)}]{ELPH}
\bibinfo{title}{{Reserach Center for Electron Photon Science (ELPH), Tohoku
  University}}, \bibinfo{year}{url=}.
  \bibinfo{note}{\url{https://www.lns.tohoku.ac.jp/en/}}.
\bibitem[{{The ATLAS IBL Collaboration}(2012)}]{IBL_collaboration_2012}
\bibinfo{author}{{The ATLAS IBL Collaboration}},
\newblock \bibinfo{title}{{Prototype ATLAS IBL modules using the FE-I4A
  front-end readout chip}},
\newblock \bibinfo{journal}{JINST} \bibinfo{volume}{7} (\bibinfo{year}{2012})
  \bibinfo{pages}{P11010--P11010}. \URLprefix
  \url{https://doi.org/10.1088/1748-0221/7/11/p11010}.
  \DOIprefix\doi{10.1088/1748-0221/7/11/p11010}.
\bibitem[{Nakamura et~al.(2015)Nakamura, Arai, Hagihara, Hanagaki, Hara, Hori,
  Hirose, Ikegami, Jinnouchi, Kamada, Kawagoe, Kohno, Motohashi, Nishimura,
  Oda, Otono, Takubo, Terada, Takashima, Tojo, Unno, Usui, Wakui, Yamaguchi,
  Yamamoto, and Yamamura}]{Nakamura2015}
\bibinfo{author}{K.~Nakamura}, \bibinfo{author}{Y.~Arai},
  \bibinfo{author}{M.~Hagihara}, \bibinfo{author}{K.~Hanagaki},
  \bibinfo{author}{K.~Hara}, \bibinfo{author}{R.~Hori},
  \bibinfo{author}{M.~Hirose}, \bibinfo{author}{Y.~Ikegami},
  \bibinfo{author}{O.~Jinnouchi}, \bibinfo{author}{S.~Kamada},
  \bibinfo{author}{K.~Kawagoe}, \bibinfo{author}{T.~Kohno},
  \bibinfo{author}{K.~Motohashi}, \bibinfo{author}{R.~Nishimura},
  \bibinfo{author}{S.~Oda}, \bibinfo{author}{H.~Otono},
  \bibinfo{author}{Y.~Takubo}, \bibinfo{author}{S.~Terada},
  \bibinfo{author}{R.~Takashima}, \bibinfo{author}{J.~Tojo},
  \bibinfo{author}{Y.~Unno}, \bibinfo{author}{J.~Usui},
  \bibinfo{author}{T.~Wakui}, \bibinfo{author}{D.~Yamaguchi},
  \bibinfo{author}{K.~Yamamoto}, \bibinfo{author}{K.~Yamamura},
\newblock \bibinfo{title}{{Irradiation and testbeam of KEK/HPK planar p-type
  pixel modules for HL-LHC}},
\newblock \bibinfo{journal}{JINST} \bibinfo{volume}{10} (\bibinfo{year}{2015})
  \bibinfo{pages}{C06008}. \URLprefix
  \url{https://iopscience.iop.org/article/10.1088/1748-0221/10/06/C06008/pdf}.
  \DOIprefix\doi{10.1088/1748-0221/10/06/C06008}, \bibinfo{note}{cited By 5}.
\bibitem[{qst(url)}]{qst}
\bibinfo{title}{{QST Cobalt irradiation facility (in Japanese)}},
  \bibinfo{year}{url=}.
  \bibinfo{note}{\url{https://www.qst.go.jp/site/taka-shisetsuka/2103.html}}.

\end{thebibliography}


\end{document}